\documentstyle[epsf,eqsecnum,aps]{revtex}

\begin{document}
\hspace{-15mm}
\vspace{-10.0mm} 

\thispagestyle{empty}
{\baselineskip-4pt
\font\yitp=cmmib10 scaled\magstep2
\font\elevenmib=cmmib10 scaled\magstep1  \skewchar\elevenmib='177
\leftline{\baselineskip20pt
\hspace{12mm} 
\vbox to0pt
   { {\yitp\hbox{Osaka \hspace{1.5mm} University} }
     {\large\sl\hbox{{Theoretical Astrophysics}} }\vss}}

\rightline{\large\baselineskip20pt\rm\vbox to20pt{
\baselineskip14pt
\hbox{OU-TAP 50}
\hbox{kucp0107}
\hbox{gr-qc/9705xxx}
\vspace{2mm}
\hbox{21 May 1997}
\hbox{version 8.2}\vss}}

\vspace{2cm} 

\begin{center}
{\Large\bf Classical Nature of the Inflaton Field } 
{\Large\bf with Self-Interaction }
\end{center}
\bigskip

\centerline{\large Takahiro Tanaka$^1$\footnote{Electronic address: 
tama@vega.ess.sci.osaka-u.ac.jp}
and Masa-aki Sakagami$^{2}$\footnote{
Electronic address: 
sakagami@phys.h.kyoto-u.ac.jp} }
\bigskip
\begin{center}
{\em $^1$Department of Earth and Space Science, 
Graduate School of Science,} 
{\em  Osaka University, Toyonaka 560, Japan}\\
{\em $^2$Cosmology Group, 
Faculty of Integrated Human Studies, 
Kyoto University, Kyoto 606-01, Japan}\\
\end{center}

\bigskip

\begin{abstract}
Taking into account the effect of self-interaction, 
the dynamics of the quantum fluctuations of the 
inflaton field with $\lambda\phi^4$ potential 
is studied in detail. 
We find that the self interaction efficiently drives 
the initial pure state into a mixed one, which can 
be understood as a statistical ensemble. 
Further, the expectation value of the squared 
field operator is found to be converted into 
the variance of this statistical ensemble without 
giving any significant change in its amplitude. 
These results verify the ansatz of the quantum-to-classical 
transition that has been assumed in the standard evaluation of 
the amplitude of the primordial 
fluctuations of the universe. 

\noindent
PACS number(s): 98.80.Cq, 04.62.+v, 05.40.+j
\end{abstract}


\section{Introduction} 
The inflationary universe scenarios explain 
satisfactorily various aspects of 
the universe, such as the homogeneity, 
isotropy and the amplitude of the primordial fluctuations 
observed in the microwave background radiation 
and in the large scale structure\cite{Linde}. 
However, the incompleteness in the 
discussion of the evaluation of the primordial fluctuations
has been pointed out in several papers as is explained below.
 
In most of inflationary universe scenarios, 
the seed of the inhomogeneity of the 
universe is traced back to the quantum fluctuations of the 
inflaton field generated by the 
accelerated expansion of the cosmic length scale. 
In this context, the amplification of 
the quantum fluctuations is characterized by a large 
amount of squeezing of the state vector. 
A difficulty in interpreting this sate vector exists 
in the fact 
that the expectation value of the linear field operator 
does not show inhomogeneities but vanishes 
while that of the squared field operator becomes 
very large. 
Usually, one interprets this quantum state as if it were 
equivalent to a statistical ensemble which has the same 
amount of variance 
that the corresponding quantum operator 
has in the sense of an expectation value 
once the scale of the fluctuations of our interest exceeds  
the Hubble horizon scale. 
Here we refer to the calculation 
based on this ad-hoc classicalization ansatz as 
``the standard calculation''.

In order to justify the standard calculation, 
the stochastic approach to the inflationary universe 
scenario was proposed by Starobinsky\cite{Starob} and 
further investigated by many authors \cite{NamSas}. 
The evolution of the order parameter that is defined as 
the expectation value of the 
spatially averaged field operator was studied. 
If the physical size of the spatial averaging is 
kept constant, the modes with a large comoving wave number 
become to contribute to the order parameter 
as the universe expands.  
The contribution from the newly added small scale modes 
affects on the dynamics of the order parameter as a random noise. 
As a consequence, the evolution of the order parameter mimics 
the Brownian motion and thus the distribution of 
the order parameter can be well approximated by a statistical 
ensemble of a classical system influenced by the stochastic noise. 

In these early studies, a free scalar field was 
investigated and little attention was paid to 
the effect of interaction.
However, the fluctuations of a free scalar 
field in a Friedmann-Robertson-Walker spacetime can 
be decomposed into a set of harmonic oscillators which have 
time dependent spring constants. 
Then all of these oscillators are decoupled with each other. 
Hence, no classicalization will be 
expected in such a decomposed degree of freedom.
It follows that the each modes, which are added to the order parameter 
during the inflation, never lose their quantum nature.
This fact implies that
the discussion of the classicalization 
in the context of the stochastic 
inflation is incomplete\cite{Habib}. 
We think that, in the true theory of the quantum-to-classical 
transition, the classical nature should be observed 
even if each decomposed mode is considered.  
In other words, we believe that 
the change of the number of modes that 
contribute to the order parameter cannot play an 
essential role in the appearance of the classical nature 
of the inflaton field. 

In this direction, recently, Lesgourgues, 
Polarski and Starobinsky\cite{lse} 
proposed a new idea to explain the quantum-to-classical 
transition. They claimed that 
the equivalence between the large squeezing 
and the classicality of the state can be 
explained if one discards the tiny contribution 
due to the decaying mode of the perturbation. 
However, a free scalar field is considered in their approach, 
and the effect of interaction was not taken 
into account manifestly. 
Thus we think that their elimination of the decaying mode is 
still artificial and will not be fully justified  
although their result seems to contain an important suggestion. 

Only recently, the importance of interaction has become 
emphasized\cite{ac,CalHu,Morikawa,Matacz}. 
As a useful tool to deal with the effect of 
interaction with the environmental degrees of freedom, 
the closed time path formalism has been developed\cite{ctp}.  
The total system is divided 
into ``the system'' and ``the environment'' so that 
the system contains the quantities of our interest while 
the environment does not. 
By using the closed time path formalism, 
one can integrate out the environmental degrees of freedom to 
obtain the effective action for the system. 
Many cosmological issues, such as 
the back reaction to the expansion rate of the universe 
due to the particle creation, 
have been investigated\cite{ac}. 

Especially, in the references\cite{Morikawa,Matacz,lla}, 
the evolution of the fluctuations of the inflaton filed 
was examined with the aid of the closed time path formalism. 
It was repeatedly stressed that 
the correct treatment of the effect of the environmental 
degrees of freedom may relax the problem of the fine tuning 
of parameters in the inflaton potential. 
Let us consider the $\lambda \phi^4$ model of the chaotic inflation. 
In the standard calculation, 
$\lambda\sim 10^{-12}$ is required in order to  
explain the observed value of the primordial 
density fluctuations\cite{Linde}. 
To the contrary, the authors in the above references 
suggested the possibility that the tuning of the 
coupling constant can be relaxed to $\lambda\sim 10^{-6}$. 
One of the main issues of the present paper is 
to cast a doubt on this statement. 

Here we take a conservative picture of the quantum-to-classical 
transition based on the paper by Joos and Zeh\cite{Zeh}. 
The basic idea is explained in Sec.~2. 
There we present two sufficient conditions 
for the system to possess the classical nature. 
Based on this picture, we investigate the $\lambda\phi^4$ model, 
which is one of the simplest models 
of the chaotic inflation\cite{Linde}. 
With the choice of parameter $\lambda\sim 10^{-12}$, 
the two conditions for the quantum-to-classical transition 
are proved to be satisfied 
and the fluctuations of the scalar field, $\delta \phi$ 
become of $O(H)$ as is predicted in the standard calculation. 
Hence, we obtain the conclusion 
that ``the standard calculation'' is justified 
in this simple model 
without any significant modification in the amplitude of 
the fluctuations as opposed to the prediction given in 
literature\cite{Morikawa,lla,Matacz}. 

This paper is organized as follows. 
In Sec.~2 we explain our picture of the quantum-to-classical 
transition and clarify in what situation the system behaves 
as a classical one. 
In Sec.~3 we explain our simple model and the basic assumptions. 
Integrating out the environmental degrees of freedom, we  
derive the effective action for the system. 
In Sec.~4 by using the closed time path formalism, 
the dynamics of the system is analyzed in a less rigorous but a 
rather intuitive manner. 
More rigorous treatment based on the quantum master equation 
for the reduced density matrix is provided in Sec.~5. 
Section 6 is devoted for summary and discussion. 

\section{decoherence measure between different worlds}

In this section, we explain our picture of the 
quantum-to-classical transition. 
At first, let us briefly summarize the standard discussion 
of decoherence based on the analysis of the reduced density matrix. 
Suppose that the whole system can be divided into two parts: one is 
the ``system'', which contains the observables of our interest, 
and the other is the ``environment'', 
which is to be integrated out to obtain the reduced density 
matrix. We set an initial condition for the whole system 
so that the density matrix is given by that of a pure state and 
the correlation between the system and the environment is absent.
Hence the state can be represented as
\begin{equation} 
\rho(t_i)  = |{t_i}\rangle_{\rm sys}\otimes|{t_i}\rangle_{\rm env}
     ~_{\rm env}\langle{t_i}|\otimes{}_{\rm sys}\langle{t_i}|. 
\end{equation}
The reduced density matrix is obtained by taking 
the partial trace over the environment,
\begin{equation}
\tilde{\rho}(x,x';t) := {\rm Tr}_{\rm env}
    ~_{\rm sys}\langle{x}|\rho(t)|{x'}\rangle_{\rm sys},
\end{equation}
where $x$ and $x'$ are the labels of a complete set of 
state vectors of the system, say, 
the eigen state of the coordinate. 
If the interaction between the system and 
the environment is absent, the loss of coherence does not take place, 
i.e. $\tilde\rho(x,x';t)$ keeps the form of 
a pure state, i.e., $\tilde\rho(x,x';t)=
 \,_{\rm sys}\langle{x}|{t}\rangle_{\rm sys}
 {\,}_{\rm sys}\langle{t}|{x'}\rangle_{\rm sys}$. 
Hence, $|\tilde\rho(x,x';t)|^2
/\tilde\rho(x,x;t)\tilde\rho(x',x';t)=1$. 
However, in the presence of interaction, 
$\tilde\rho(x,x';t)$ no longer keeps the form of 
a pure state. 
If $|\tilde\rho(x,x';t)|^2
/\tilde\rho(x,x;t)\tilde\rho(x',x';t)$ 
becomes quite small for 
$x\ne x'$, we may say that the coherence 
between different states labeled by $x$ and $x'$ disappears. 
If $x$ is a continuous parameterization of state vectors, 
we can determine the typical scale $\Delta x$ such that 
the coherence between the states labeled by $x$ and $x'$ 
is lost when $|x-x'|>\Delta x$. 
As long as we observe the system with a resolution coarser 
than $\Delta x$, we may think that the different states 
have no interference between them and so they can be 
recognized as independent different worlds. 
However, in the above discussion, 
the stability of the state through the time evolution 
was not taken into account. 
The dynamics of the system itself and the effects from 
the environment cause the broadening of the wave function 
of the system in general. If there is a large amount of 
broadening, it would be difficult to 
interpret that the state evolves into a statistical 
ensemble of many different ``classical worlds''.

Here we propose to 
take more conservative picture of the quantum-to-classical 
transition in this paper. 
As mentioned in Introduction, 
the basic idea is taken from the paper 
by Joos and Zeh\cite{Zeh}. 
We restrict our attention to the case in which 
the interaction term between the system and the environment 
does not contain the momentum variable of the system. 
We suppose that the sufficient conditions in order 
for the system to possess the classical nature are the 
following two. 

The first condition is that 
the total system has a set of wave packets which 
have a sufficiently peaky probability distribution 
in comparison with the accuracy of our measurement  
through the whole duration we consider.
Say, we label the wave packets by their initial peak position, $s$,  
as $|{s;t}\rangle$. 
Since the interaction term does not contain 
the momentum variable, in an approximate sense 
each wave packet will be written by the direct 
product as 
\begin{equation}
 |{s;t}\rangle=|{s;t}\rangle_{\rm sys}\otimes |{s;t}\rangle_{\rm env}. 
\end{equation}
Here we should note that the state of the environment 
$|{s;t}\rangle_{\rm env}$ is also labeled by $s$, since 
the interaction causes the correlation between the system 
and the environment in the course of their time 
evolution\cite{simple}. 
Strictly speaking, what we require here is that 
not $|{s;t}\rangle$ but $|{s;t}\rangle_{\rm sys}$ has a sharp peak 
and that the peak is stable against the evolution. 
Furthermore, we require that 
this set of wave packets is complete enough that 
the initial state, $|{\Psi}\rangle$, can be decomposed into 
a quantum mechanical superposition of these wave packets, i.e.,
\begin{equation} 
 |{\Psi}\rangle=\sum_s c_s |{s;t}\rangle. 
\end{equation}
This means that each wave packet is sufficiently peaky and stable 
to be recognized as distinguishable ``world''.
We refer to this condition as 
``the classicality of the dynamics of the system''. 

The second condition is that the coherence between the different 
wave packets becomes lost swiftly. 
We can say that this condition is the one that was 
roughly discussed at the beginning of this section. 
We refer to this condition as ``the decoherence between different worlds''. 
For the total system that satisfies the above mentioned 
first condition, the density matrix at time $t$ will be given by 
\begin{equation}
\rho(t) = \sum_{s,s'}c_s c^*_{s'} |{s;t}\rangle\langle{s';t}|, 
\end{equation}
The partial trace over the environment gives the reduced density matrix,
\begin{equation}
\tilde{\rho}(t) = 
\sum_{s,s'}c_s c^*_{s'} C(s,s';t) |{s;t}\rangle_{\rm sys}~_{\rm sys}
   \langle{s';t}| 
=: \sum_{s,s'} \tilde{\rho}_{s,s'}(t) ,
\label{redden}
\end{equation}
where we defined the partial reduced 
density matrix $\tilde\rho_{s,s'}(t)$ and 
$ C(s,s';t) $ is given by
\begin{equation}
 C(s,s';t):={}_{\rm env}\langle s;t\vert s';t\rangle_{\rm env}. 
\label{Cdef}
\end{equation}
If we assume that the initial state is given by a direct 
product of the state of the system and that of the environment, 
we find $C(s,s';t_i)=1$. 
If the interaction between the system and 
the environment is absent, the loss of coherence does not take place, 
i.e. $C(s,s';t)$ stays time independent constant for all $s$ and $s'$. 
However, in the presence of interaction, 
$\vert s;t \rangle_{\rm env}$ and 
$\vert s';t \rangle_{\rm env}$ evolve differently and hence 
$C(s,s';t)$ no longer stay constant for $s\ne s'$, while 
$C(s,s;t)\equiv 1$. 
If $|C(s,s';t)|$ becomes quite small for $s\ne s'$, 
we can say that the diagonalization of $\tilde \rho$ has 
been occurred. 
Thus the quantity $|C(s,s';t)|$ 
characterizes the degree of decoherence between two 
different worlds, if the evolution of the wave packets is not seriously 
affected by the environment.

Here we note that 
\begin{equation}
 |c_s c^{*}_{s'} C(s,s';t)|^2={\rm Tr} \left(\tilde\rho_{s,s'}(t) 
  \tilde\rho^{\dag}_{s,s'}(t)\right), 
\label{Csq}
\end{equation}
In a strict sense, $\vert s;t \rangle$ can not be written by 
a direct product as 
$\vert s;t \rangle_{\rm sys} \otimes \vert s;t \rangle_{\rm env}$. 
So the definition of $C(s,s';t)$ given in (\ref{Cdef}) is 
not well defined. 
However, the expression in the right hand side of Eq.~(\ref{Csq}) 
makes sense at any time and is 
expected to give the measure of the decoherence between different worlds. 
 
As a more convenient alternative of the measure of decoherence, 
we propose to use 
\begin{equation}
R  =  {{\rm max}|\tilde\rho_{s,s'}(t)| 
  \over {\rm max}|\tilde\rho_{s,s'}(t_i)|}. 
\label{defdeco1}
\end{equation}
If $\tilde\rho_{s,s'}(t)$ takes the Gaussian form, 
$R$ equals to $|C(s,s';t)|$ besides the small correction 
due to the determinant factor 
that arises from the Gaussian integral. 
If $R$ becomes very small for $s\ne s'$, we recognize that 
the decoherence between different worlds is achieved. 

In some sense, our picture is that of the third quantization of 
the universe\cite{tq} or of the decoherence history\cite{dh}. 
Our two requirements for the quantum-to-classical transition 
may be too strong\cite{qdcc}. 
But, for the present purpose, we do not have to relax 
these conditions.

\section{Simple model and the effective action}

We consider the following simple model of inflation 
consisted of a single real scalar field 
using an approximation similar to the one that introduced 
by Matacz\cite{Matacz}, although the details are significantly modified. 
We assume that the space time can be approximated by 
a spatially flat de Sitter space, 
\begin{equation}
 ds^2=dt^2-a^2(t)d^2x, 
\end{equation}
with 
\begin{equation}
  a(t)={1\over H} e^{Ht},
\end{equation}
and that the Lagrangian of the inflaton field is given by 
\begin{equation}
 S=\int_{t_{i}}^{t}ds~a^3(s)\int_{\Omega}d^3 x \left[
  {1\over 2} \left({d\Phi(x,s)\over ds}\right)^2 
  -{\left(\nabla\Phi(x,s)\right)^2\over 2 a^2(s)}
  -V\left(\Phi(x,s)\right)\right], 
\end{equation} 
where $\Omega$ is a finite comoving volume corresponding to 
the scale of the fluctuations of our interest and we assumed 
that the effect from outside of this volume can be 
neglected. For simplicity, we choose $\Omega$ as a cube with 
$0\le x,y,z \le L$. 
The lower boundary of the time integration, $t_{i}$, is the time 
at which some appropriate initial condition is set. 

We focus on the dynamics of a spatially averaged field 
in this volume, 
\begin{equation}
 \phi(s):={1\over \Omega}\int_{\Omega} d^3x\, \Phi(x,s). 
\end{equation}
We stress that this averaging is performed only on 
a finite comoving volume, i.e., a part of the time-constant 
spatial surface. Hence $\phi(s)$ does not represent the 
homogeneous part of the field $\Phi(x,s)$ but it represents 
the fluctuation of scale $L$. 
Although it is not essential but, 
for definiteness, we set the periodic boundary condition on $\Phi(x,s)$. 
Then $\Phi(x,s)$ is decomposed as 
\begin{eqnarray}
 \Phi(x,s) & = & \phi(s)+\psi(x,s)\cr
  &=& \phi(s)+\sqrt{2\over \Omega}\sum_k\left[
  q_{\bf k}^{+} \cos {\bf k}\cdot {\bf x} 
  +q_{\bf k}^{-} \sin {\bf k}\cdot {\bf x}\right], 
\end{eqnarray}
where $\displaystyle {\bf k}={2\pi\over L}(i,j,k)$ is 
a non-vanishing vector with 
integer, $i,j$, and non-negative integer, $k$. 
Assuming that the potential can be approximated by 
\begin{equation}
 V(\Phi)\sim V(\phi)+\psi V'(\phi)+{\psi^2\over 2}V''(\phi), 
\end{equation} 
the action reduces to 
\begin{eqnarray} 
 S[\phi,q] & = & \Omega\int_{t_{i}}^t ds~ a^3(s)\left[ 
 {1\over 2} \dot\phi^2(s)-V\left(\phi(s)\right)\right] 
\cr
 && -{1\over2} \sum_{\sigma}^{\pm} \sum_{\bf k} 
   \int_{t_{i}}^t ds~ a^3(s)V''\left(\phi(s)\right)\left(
    q_{\bf k}^{\sigma}\right)^2
\cr
 && -{1\over2} \sum_{\sigma}^{\pm} \sum_{\bf k} 
   \int_{t_{i}}^t ds~ a^3(s) \left[
    \left(\dot q_{\bf k}^{\sigma}\right)^2
     -{{\bf k}^2\over a^2(s)} 
      \left(q_{\bf k}^{\sigma}\right)^2\right]
\cr
  & =: &  S_{\rm sys}[\phi] +S_{\rm int}[\phi,q]+S_{\rm env}[q],
\label{actiondec}
\end{eqnarray}
where dot $~\cdot{}~$ means a derivative with respect to $t$. 
Expecting that the interaction term is small, i.e., 
$V''(\phi(s))$ is so, 
we consider the perturbative expansion with respect to it. 
As seen from the notation introduced in Eq.~(\ref{actiondec}), 
the spatially averaged field is considered as the system and 
the other short wave length modes are the environment. 

As was performed in Ref.\cite{Matacz}, 
we calculate the reduced density 
matrix for $\phi$ integrating over the environmental degrees of 
freedom, $q_{\bf k}^{\sigma}$. 
Here we assume that the density matrix of 
the total system is initially represented by the direct product as 
\begin{equation}
 \rho_i=\tilde \rho(\phi_i,\phi'_i;t_{i})\otimes 
        \prod_{\sigma}^{\pm}\prod_{\bf k}
       \rho_{\bf k}^{\sigma}
     (q_{{\bf k}i}^{\sigma},{q_{{\bf k}i}^{\sigma}}';t_{i}). 
\end{equation}
In order to define the initial quantum state of $q_{\bf k}^{\sigma}$, 
we suppose that the interaction is switched off before $t=t_{i}$ and 
we set the quantum state of $q_{\bf k}^{\sigma}$ by 
\begin{equation}
  a_{\bf k}^{\sigma} \vert 0\rangle =0, 
\end{equation}
where the annihilation operator $a_{\bf k}^{\sigma}$ 
is defined by 
the decomposition of the field operator, 
\begin{equation}
  \hat q_{\bf k}^{\sigma}=a_{\bf k}^{\sigma} u_{\bf k}^{\sigma}+
                     a_{\bf k}^{\dag\sigma} u_{\bf k}^{\sigma *},
\end{equation}
and the positive frequency function, $u_{\bf k}^{\sigma}$, is taken as 
\begin{equation}
 u_{\bf k}^{\sigma}=u_0
 :={1\over \sqrt{2k}}
 {e^{-ik\eta}\over a(t)}\left(1-{i\over k\eta}\right),
\end{equation} 
before $t=t_{i}$. 
Here we introduced the conformal time coordinate by 
$\eta:=-e^{-Ht}$. 
Then the reduced density matrix is calculated to the second order in 
$S_{\rm int}[\phi,q]$ as \cite{Matacz}
\begin{eqnarray}
 \tilde\rho(\phi,\phi';t) & := & \int dq~ \rho(\phi,\phi',q,q';t)
\cr
 & = &\int d\phi_i \int d\phi'_i
      \int_{\phi_i}^{\phi}{\cal D}\phi 
      \int_{\phi'_i}^{\phi'}{\cal D}\phi' 
      \exp \left[i\left\{S_{\rm sys}[\phi]-S_{\rm sys}[\phi']\right\}
        +iS_{\rm IF}[\phi,\phi']\right]
     \tilde\rho(\phi_i,\phi'_i;t),
\label{effective}
\end{eqnarray}
and 
\begin{eqnarray}
iS_{\rm IF}[\phi,\phi']&=&-i\int_{t_{i}}^{t} ds~ \Delta(s) f(s) 
       +i\int_{t_{i}}^{t} ds \int_{t_{i}}^{s} ds'~ 
          \Delta(s) \Sigma(s') \mu(s,s') 
\cr &&\quad
       -\int_{t_{i}}^{t} ds \int_{t_{i}}^{s} ds'~ 
          \Delta(s) \Delta(s') \nu(s,s'),
\end{eqnarray} 
where 
\begin{equation}
 \Delta(s)= V''(\phi)-V''(\phi'),\quad 
 \Sigma(s)= {1\over 2} \left(V''(\phi)+V''(\phi')\right),
\end{equation}
and 
\begin{eqnarray}
 f(s) & = & { a(s)^3 \over 2}\sum_{\sigma}^{\pm} \sum_{\bf k} 
        u_0^{*}(s) u_0(s), 
\cr
 \mu(s,s') & = & -i{a(s)^3 a(s')^3\over 2} 
        \sum_{\sigma}^{\pm} \sum_{\bf k} 
        \left([u_0^{*}(s)]^2 [u_0(s')]^2 -
          [u_0(s)]^2 [u_0^{*}(s')]^2 \right),
\cr
 \nu(s,s') & = & {a(s)^3 a(s')^3\over 4} 
        \sum_{\sigma}^{\pm} \sum_{\bf k} 
        \left([u_0^{*}(s)]^2 [u_0(s')]^2 + 
          [u_0(s)]^2 [u_0^{*}(s')]^2 \right).
\end{eqnarray}

In order to evaluate the summation over the discrete modes, 
we replace it with the integral as 
\begin{equation}
  \sum_{\sigma}^{\pm} \sum_{\bf k} \rightarrow 
  {\Omega\over 2\pi^2}\int_{k_{min}}^{\infty}k^2~ dk, 
\end{equation}
where we take $k_{min}$ as a constant of $O(2\pi/L)$. 
Since the proper length scale, $a(t) L$, of the fluctuations
which contribute to the formation of the large scale structure or 
the anisotropies of the cosmic microwave background radiation 
becomes much larger than the Hubble scale, $H^{-1}$ at the end 
of the inflation era. 
Therefore for simplicity we concentrate on the case in which 
\begin{equation}
 a(t)H=-\eta^{-1}\gg k_{min},
\label{etacondi}
\end{equation}
is satisfied. 
Thus we set the initial condition for the reduced density 
matrix at a time after the length scale of the fluctuations 
of our interest exceeds the Hubble horizon scale. 
Or equivalently, we assume that 
the evolution is free from the effect 
of the environmental degrees of freedom until 
$t=t_{i}$. 
Under this condition, we evaluate 
the functions $f(s)$, $\mu(s,s')$ and $\nu(s,s')$ approximately. 
The evaluation of $f(s)$, $\mu(s,s')$ and $\nu(s,s')$ is rather 
complicated. The details of computation are given in appendix A. 
As shown there, $f$ and $\mu$-terms contain the 
ultraviolet divergences that require renormalization. 
After subtraction of these divergences, keeping only the leading terms, 
we obtain 
\begin{eqnarray}
 f(s)
    & \sim & {a^3(s)\Omega H^2\over 8\pi^2}
    \log\left({H\over p_{min}(s)}\right),
\cr
 \nu(s,s') 
   & \sim & {a^3(s)\Omega H^4\over 48\pi^2 p_{min}^3(s)}e^{-3H(s-s')}, 
\label{fmunu}
\end{eqnarray}
where we introduced $p_{min}(s):=a^{-1}(s) k_{min}$ and 
$p_{max}(s):=a^{-1}(s) k_{max}$. 
As shown in appendix A, in order to subtract the divergent portion 
in the $\mu$-term of $S_{\rm IF}$, 
a integration by part with respect to $s'$ is necessary. 
Then $\mu$-term is evaluated as 
\begin{equation}
S_{\rm IF}^{(\mu)} = \int_{t_{i}}^{t} ds\Delta(s) \int_{t_{i}}^s 
  ds' \left[2\left(\mu_i(s,s') +\mu_a(s,s')\right)\Sigma(s')+ 
  \mu_b(s,s') \dot \Sigma(s')\right],  
\label{SIF}
\end{equation}
where 
\begin{eqnarray}
 \mu_i(s,s') & \sim & -{a^3(s)\Omega\over 16\pi^2}\log
    \left({p_{min}(s)\over H}\right) \delta(s-s'), \cr
 \mu_a(s,s') & \sim & -{a^3(s)\Omega H\over 16\pi^2}\log
    \left(k_{min}(\eta-\eta')\right), \cr
 \mu_b(s,s') & \sim & {a(s)a^2(s')\Omega\over 16\pi^2}\log
    \left(k_{min}(\eta-\eta')\right). 
\label{musdef}
\end{eqnarray}
Precisely speaking, the approximated values of $\mu$s 
are different from the true value by a factor of order unity. 
Thus $16\pi^2$ can be replaced by, say, $24\pi^2$. 
However, these errors do not change the discussion 
given below in this paper 
because we just show that the effect of these terms 
is small and can be neglected.

\section{an Intuitive interpretation of the effective action}

Here we give a less rigorous but a rather intuitive 
analysis of the evolution of the averaged field, $\phi$, 
in the model with $\displaystyle V={\lambda\Phi^4\over 4!}$. 
We defer a more rigorous treatment to the next section. 

The effective action is rewritten in terms of 
\begin{equation}
  \phi_{+}={\phi+\phi'\over 2}, \quad
  \varphi_{\Delta}=\phi-\phi',
\end{equation}
as
\begin{eqnarray}
S[\phi,\phi']=&&\Omega\int_{t_{i}}^{t}ds~
  a^3(s)\left\{\dot\phi_{+}(s)\dot\varphi_{\Delta}(s) 
        -V'(\phi(s))\varphi_{\Delta}\right\}
\cr
 &&-\lambda\int_{t_{i}}^{t} ds~\phi_{+}(s)
         \varphi_{\Delta}(s)f(s)
   +i \lambda^2 \int_{t_{i}}^{t} ds \phi_{+}(s)
         \varphi_{\Delta}(s) \int_{t_{i}}^{s} ds'~
         \nu(s,s')\phi_{+}(s')
         \varphi_{\Delta}(s') + S^{(\mu)}[\phi,\phi'],
\end{eqnarray}
and 
\begin{equation}
S^{(\mu)}[\phi,\phi']= \lambda^2\int_{t_{i}}^{t} ds\phi_{+}(s)
       \varphi_{\Delta}(s) \int_{t_{i}}^s 
  ds' \left[\left(\mu_i(s,s') +\mu_a(s,s')\right)\phi_{+}^2(s')+ 
  \mu_b(s,s') \phi_{+}(s')\dot\phi_{+}(s') \right],  
\end{equation}
where the cubic or higher terms with respect to 
$\varphi_{\Delta}$ are neglected.

Since $\nu(s,s')$ decays fast as $s-s'$ becomes large, 
here we approximate it as 
\begin{equation}
 \nu(s,s')\sim \delta(s-s') {a^6(s)\Omega^2 \Xi^2(s)}/\phi_+^2(s),
\end{equation}
where
\begin{equation}
\Xi(s):={\phi_+ (s)\over a^3(s)\Omega}
   \left[\int_{t_{i}}^{s} ds'\,2\nu(s,s')\right]^{1/2}
    \sim \sqrt{H^3\over 72 \alpha' \pi^2} 
     \phi_{+}(s),
\end{equation}
with $\alpha':=\Omega k_{min}^3$. 
Then the $\nu$-term in the action reduces to 
\begin{equation}
S_{\nu}[\phi,\phi']
    =i\lambda^2 \Omega^2\int_{t_{i}}^t ds~
   a^6(s)\Xi^2(s)
         {\varphi^2_{\Delta}(s)\over 2}.
\label{action2}
\end{equation}
To manage the effect of the $\nu$-term, we introduce 
an augsiliary field, $\xi$, 
which represents the Gaussian white noise 
\begin{equation}
 \langle\xi(s)\xi(s')\rangle_{\rm EA}=\delta(s-s'), 
\end{equation}
where the subscript EA stands for the ensemble average. 
Further we introduce the action with noise by 
\begin{eqnarray}
S_{\xi}[\phi,\phi']=&&\Omega\int_{t_{i}}^{t}ds~
  a^3(s)\left\{\dot\phi_{+}(s)\dot\varphi_{\Delta}(s) 
        -V'(\phi(s))\varphi_{\Delta}\right\} 
\cr
 &&-\lambda\int_{t_{i}}^{t} ds~\left(\phi_{+}(s)f(s)
         -a^3(s)\Omega \Xi(s)\xi(s)\right) \varphi_{\Delta}(s)
     +S^{(\mu)}[\phi,\phi'].
\end{eqnarray}
Then from the fact that 
\begin{equation}
 \exp\left(iS[\phi,\phi']\right)
  =\left\langle\exp\left(iS_{\xi}[\phi,\phi']\right)
  \right\rangle_{\rm EA}, 
\end{equation}
we can expect that the action with noise, 
$S_{\xi}$, 
gives the evolution of the field including the 
effect of the fluctuations induced by the environment 
through the $\nu$-term. 
Taking the variation 
of $S_{\xi}$ with respect to 
$\varphi_{\Delta}$, 
we obtain the Heisenberg equation 
for the operator $\hat\phi_{+}(t)$. 
Sandwiching thus obtained Heisenberg equation in between the ``bra'' 
and ``cket'' vectors 
of the initial state, the equation for the expectation 
value becomes 
\begin{eqnarray}
  {d^2\over dt^2}\langle{\hat\phi}_{+}(t)\rangle
   &+&3H{d\over dt}\langle{\hat\phi}_{+}(t)\rangle
   +{\lambda\over 6}\langle(\hat\phi^3_{+}(t))\rangle
   +{\lambda f(t)\over a^3(t)\Omega}
     \langle\hat\phi_+(t)\rangle\cr
&&   -{\lambda^2\over a^3(t)\Omega}
   \int_{t_{i}}^{t} ds~
   \left[\left(\mu_i(t,s)+\mu_a(t,s)\right)
   \langle\hat\phi_+(t)\hat\phi_+^2(s)\rangle 
   +\mu_b(t,s)
   \langle\hat\phi_+(t)\hat\phi_+(s)\dot{\hat\phi}_+(s)
    \rangle  \right]
   =\Xi(t)\xi(t). 
\end{eqnarray}

If we set 
\begin{equation}
 \hat\phi_+ =\langle\hat\phi_+\rangle+\hat\varphi,
\end{equation}
$\langle\hat\varphi\rangle=0$ follows by construction.
Thus we find 
\begin{equation}
 \langle\hat\phi_+^3(t)\rangle
 =\langle\hat\phi_+(t)\rangle^3 
 +3\langle\hat\phi_+(t)\rangle \langle\hat\varphi^2(t)\rangle
 +\langle\hat\varphi(t)\rangle^3. 
\end{equation}
Assuming that the effect of the quantum fluctuations 
is negligible, i.e., $\langle\hat\varphi^2(t)\rangle \ll 
\langle\hat\phi_+(t)\rangle^2$, 
we obtain 
\begin{eqnarray}
  {d^2\over dt^2}\langle{\hat\phi}_{+}(t)\rangle
   &+&3H{d\over dt}\langle{\hat\phi}_{+}(t)\rangle
   +{\lambda\over 6}\langle\hat\phi_{+}(t)\rangle^3
   +{\lambda f(t)\over a^3(t)\Omega}
     \langle\hat\phi_+(t)\rangle \cr
   &-& {\lambda^2\over a^3(t)\Omega}
   \langle\hat\phi_+(t)\rangle \int_{t_{i}}^{t} ds~
   \left[\left(\mu_i(t,s)+\mu_a(t,s)\right)
   \langle\hat\phi_+(s)\rangle^2 
   +\mu_b(t,s)
   \langle\hat\phi_+(s)\rangle 
   {d\langle{\hat\phi_+}(s)\rangle \over ds}
     \right]
   =\Xi(t)\xi(t). 
\label{Heqphi}
\end{eqnarray}
In the rest of this section, 
we simply use $\phi(t)$ instead of  
$\langle\phi_+(t)\rangle$. 
The $f$- and $\mu$-terms give the correction to the 
evolution of $\phi$ in a deterministic 
manner while the effect of $\nu$-term gives 
a stochastic force. 

Now we show that the effect of 
the $f$- and $\mu$-terms are negligible under the 
present condition. 
The effect of $f$-term is just to modify the potential. 
Thus we compare the force due to the $f$-term:
\begin{equation}
 {\lambda f(t) \phi(t)\over a^3(t)\Omega}\sim
 {1\over 8\pi^2}\lambda \phi(t) H^2\log\left(H\over p_{min}(t)\right),
\label{eq413}
\end{equation}
with that due to the bare potential (the third term 
in Eq.~(\ref{Heqphi})): 
\begin{equation}
 V'\left(\phi(t)\right)=
  {\lambda \phi^3(t)\over 6}.
\label{comp}
\end{equation}
Hence, the contribution from $f$-term 
can be neglected if $\phi^2\gg H^2\log(H/p_{min}(t))$. 
Since in the inflationary universe scenario 
the typical value of $\phi$ at the time 
when the comoving scale of the fluctuations of our interest
crosses the Hubble horizon scale 
during inflation is known to become the order of the 
Planck scale, $m_{pl}$. 
Therefore the above inequality holds. 
Thus we conclude that the $f$-term can be neglected. 

The effect of $\mu$-term is not a simple change of the 
potential but has a hereditary one. 
Again the force coming from the $\mu_i$ and $\mu_a$-terms 
is roughly evaluated as 
\begin{eqnarray}
\Biggl\vert-{\lambda^2\over a^3(t)\Omega} \phi(t)
   \int_{t_{i}}^{t} &ds& ~
   \left(\mu(t,s)_i+\mu_a(t,s)\right)\phi^2(s)\Biggl\vert \cr
 &\sim & \left\vert{\lambda^2\over 16\pi^2}\phi^3(t) 
   \int_{t_{i}}^{t} ds'~ \left[\log\left({p_{min}(s')\over H}\right)
   \delta(t-s')
   +H\log\left(k_{min}(\eta_t-\eta')\right)\right]\right\vert 
\cr & \alt & 
   {\lambda^2\over 16\pi^2} \phi^3 (t) \left(H\Delta t\right)^2. 
\label{eq415}
\end{eqnarray}
where $\eta_t$ is the conformal time corresponding to 
the cosmological time $t$ and $\Delta t$ is the maximum value 
that $t-t_i$ takes.  
Here we neglected the time-dependence of $\phi(s)$ 
because the slow rolling condition is expected to 
be satisfied. 
The details of evaluation of the integral is shown in appendix B. 
Equation (\ref{eq415}) is to be compared with Eq.~(\ref{comp}). 
Then we find that the $\mu_i$ and $\mu_a$-terms 
can be neglected as long as the condition 
$\displaystyle 1\gg {\lambda\over 16\pi^2} \left(H\Delta t\right)^2
$ holds. 
This condition is satisfied for 
typical values of the model parameters such as 
$\lambda\sim 10^{-12}$ and $H\Delta t\sim 60$. 

We turn to the contribution from $\mu_b$. 
In the same way, it is evaluated as 
\begin{eqnarray}
\left\vert -{\lambda^2\over a^3(t)\Omega} \phi(t)
   \int_{t_{i}}^{t} ds~
   \mu_b(t,s)\phi(s)\dot\phi(s)\right\vert
 &\sim & 
  \left\vert{\lambda^2\over 16\pi^2}\phi^2(t)\dot\phi(t) 
   \int_{t_{i}}^{t} ds'~ {\eta_t^2\over {\eta'}^2}
   \log\left(k_{min}(\eta_t-\eta')\right)\right\vert 
\cr & \alt & 
   {\lambda^2\over 32\pi^2} \phi^2(t){\dot\phi(t)}\Delta t. 
\label{eq415p}
\end{eqnarray}
Also, the details of calculation are provided in appendix B. 
This term should be compared with the friction term due to 
the cosmic expansion (the second term in Eq.~(\ref{Heqphi})):
\begin{equation}
3H\dot\phi(t). 
\end{equation}
Then the ratio of these two terms is evaluated as  
$\displaystyle 
{\lambda\over 48\pi^2}\left({v\over H^2}\right)
    H\Delta t$, and is 
found to be small. Here we introduced a constant, $v$, as 
a typical value of the inflaton mass squared: 
$v\sim \lambda\phi(t)^2/2$. 
For $\lambda \phi^4$ model, typical value for $v$ is 
given by $v/H^2\sim1/100$. 
Thus we conclude that $\mu_b$-term can be also neglected. 
Thus we concentrate on the effect of the $\nu$-term 
neglecting $f$- and $\mu$-terms in the rest of this section. 

Under the condition that the fluctuation, 
$\delta\phi(t):=\phi(t)-\phi(t)|_{\xi\equiv 0}$ 
caused by $\xi(t)$ is small, 
the above equation reduces 
\begin{equation}
 \delta\ddot\phi(t)+3H\delta\dot\phi(t)
   +{V''(\phi(t))}\delta\phi(t)
   =\lambda{\Xi(t)}\xi(t). 
\end{equation}
Approximating $V''(\phi(t))$ and $\Xi(t)$ by constants $v$ 
and $\Xi$, respectively, 
the equation can be solved as 
\begin{equation}
\delta\phi(t)=-{\lambda\Xi\over D}\left[
    \int_{t_{i}}^t ds~e^{-\lambda_1 (t-s)}
    \xi(s)
    -\int_{t_{i}}^t ds~e^{-\lambda_2 (t-s)}
    \xi(s)\right],
\end{equation}
where 
\begin{eqnarray}
 && \lambda_1={3H+D\over 2}, \quad \lambda_2={3H-D\over 2},\cr
 && D =\sqrt{9H^2-4v}. 
\label{lambdaDef}
\end{eqnarray}
Since $v$ is much smaller than $H^2$ 
for $\lambda \phi^4$ model, we can approximate 
as $\lambda_1\sim3H$ and $\lambda_2\sim v/3H$. 
Then the fluctuation caused by this stochastic field $\xi(t)$ is 
evaluated as 
\begin{eqnarray}
 \langle(\delta\phi)^2\rangle_{\rm EA} 
   & \sim & {\lambda^2\Xi^2\over 2D^2} \left({D^2\over 3Hv}-
     {1\over \lambda_2}e^{-2{\lambda_2}(t-t_{i})}\right)
\cr 
   & \sim & {\lambda\over 432\alpha'\pi^2} 
   \left(1-e^{-2\lambda_2 (t-t_{i})}\right) H^2,  
\label{EAdphi}
\end{eqnarray}
which means that the $\nu$-term broadens 
the peak width of each wave packet as much as 
$\langle(\delta\phi)^2\rangle_{\rm EA}$. Hence, the effect 
is negligible small as long as the width of the 
packet, $(\delta\phi)_{\rm WP}^2$, is much larger than 
$\langle(\delta\phi)^2\rangle_{\rm EA}$.
In the references\cite{Morikawa,Matacz}, 
this quantity $\rangle(\delta\phi)^2_{\rm EA}\langle$ 
was interpreted as the real fluctuation 
which is expected to become classical. 
Thus our interpretation is totally different from theirs. 
Here we do not further discuss this issue.
A rigorous justification of our interpretation is 
provided in the next section.

The main effect of the $\nu$-term is to 
reduce the off-diagonal elements of the density 
matrix and brings the quantum state into a decohered one. 
In the above, we have shown that 
the effect of the environment is ineffective on the evolution 
of the trajectory of the peak, $\phi_{\psi}(t)$, of a wave packet, 
$\Psi_{\psi}(\phi,t)$, where $\psi$ is 
a label to distinguish different wave packets.  
Here we use the initial peak position as the label of 
wave packets, i.e., $\psi=\phi_{\psi}(t_{i})$. 

We set the initial condition at a time, $t_{i}$, 
sufficiently after the scale of the fluctuations of our interest, $L$,  
crossed the Hubble horizon, namely $a(t_{i}) L> H^{-1}$. 
Assuming that the quantum state is not affected much by 
the environment before $t=t_{i}$, 
we take the initial quantum state of 
the system as a pure state 
represented by a squeezed vacuum state 
which has a large variance 
\begin{equation}
  (\delta\phi)^2_{\rm QF}\sim H^2, 
\end{equation}
which corresponds to a natural vacuum state in 
the de Sitter space such as the Euclidean vacuum state. 
We decompose the initial wave function into 
the superposition of the wave packets, $\Psi_{\psi}(\phi,t)$.  
These wave packets are supposed to 
have a sharp peak at $\phi_{\psi}(t)$ with the width, 
$(\delta\phi)_{\rm WP}$, that is much smaller than $(\delta\phi)_{\rm QF}
\sim H$. 
We write the reduced density matrix at initial time as  
\begin{equation} 
 \tilde\rho [\phi,\phi';t_{i}]= \int d\psi \int d\psi' 
   C(\psi) \Psi_{\psi}(\phi,t_{i}) C^{*}(\psi') 
   \Psi^{*}_{\psi'}(\phi',t_{i}).
\label{initial}
\end{equation}
If the condition, $(\delta\phi)^2_{\rm QF}\gg
   \langle(\delta\phi)^2\rangle_{\rm EA}$, 
is not satisfied, this decomposition is of no use 
because we cannot construct 
the wave packets that do not lose their shape as time passes. 
(Later we find that there is another restriction related 
with the uncertainty relation.) 
Looking at Eq.~(\ref{action2}), 
the evolution of this state under the influence of the 
environment will be approximately given by 
\begin{equation} 
 \tilde\rho [\phi,\phi';t] \sim \int d\psi \int d\psi' 
   C(\psi) \Psi_{\psi}(\phi,t) C^{*}(\psi') 
    \Psi^{*}_{\psi}(\phi',t) 
   \exp\left(-\lambda^2 \Omega^2 a^6(t) \Xi^2
        {(\psi-\psi')^2\over 12H}\right), 
\label{decrho}
\end{equation}
where we thought of $\Xi(t)$ and $\phi_{\psi}(t)-\phi_{\psi'}(t)$ 
as constants $\Xi$ and $\psi-\psi'$, respectively. 
The latter replacement will be justified because 
$\dot\phi_{\psi}$ is nearly constant when 
the slow rolling condition is satisfied. 
The equation (\ref{decrho}) indicates 
that the off-diagonal elements are exponentially suppressed when 
\begin{equation} 
(\psi-\psi')^2\gg (\delta\phi)_{\rm dec}^2 
  ={6H\over \lambda^2 \Omega^2 a^6(t) \Xi^2}
  ={216\pi^2 \over \alpha'}\left({p_{min}(t)\over H}\right)^6 
    \left({\lambda v\over H^2}\right)^{-1} H^2. 
\label{phidec}
\end{equation}
The factor $(\lambda v/H^2)^{-1}$ is 
a large number typically of $O(10^{14})$, 
but $(p_{min}(t)/ H)^6$ becomes extremely small as 
$e^{-6H(t-t_{i})}\sim e^{-360}$. 
Hence, $(\delta\phi)^2_{\rm QF}\gg 
(\delta\phi)^2_{\rm dec}$. 
Therefore even if we require that the wave packets have 
a peak that is sharp enough to satisfy 
$(\delta\phi)^2_{\rm WP}\ll (\delta\phi)^2_{\rm QF}$, 
it is still possible to choose $(\delta\phi)^2_{\rm WP}$ so as 
to satisfy $(\delta\phi)^2_{\rm WP} \gg (\delta\phi)_{\rm dec}^2$. 
So, if the width of the wave packets, i.e., 
the coarse graining scale of our view, is appropriately chosen, 
they lose the quantum coherence with each other during the inflation. 
Thus we conclude that the $\nu$-term leads  
the quantum state of the system effectively into a decohered one, 
which can be recognized as a statistical ensemble of the states 
represented by wave packets with a sufficiently sharp peak, 
without any significant distortion of the shape or the peak position 
of each wave packet. 

Before closing this section 
we must mention the effect related with the uncertainty 
relation. 
Here we decomposed 
the initial quantum state of the system 
into a superposition of wave packets with a small variance with 
respect to the variables in configuration space, $\phi$. 
We write the width of the wave packet, $(\delta\phi)_{\rm WP}$ as 
$1/\sqrt{\Gamma}$ for the later convenience. 
According to the uncertainty principle, 
the small variance in $\phi$ necessarily indicates 
the existence of a large variance in its 
conjugate variable, $\Omega a^3(t)\dot \phi(t)$.  
This variance may induce a large effect on the succeeding 
evolution. 
The possible presence of the effect of this kind was 
first pointed out by Matacz\cite{dwp}. 
The induced variance in $(\delta\phi)^2$ will become of 
$O\left(\left[\int_{t_{i}}^{t}ds\,\delta\dot\phi(s)\right]^2\right)$.
This will be evaluated by using the uncertainty relation, 
$\Omega a^3(t)(\delta\dot\phi(t))_{\rm UR}
 (\delta\phi(t))_{\rm WP}\sim 1$, 
as 
\begin{equation}
  (\delta\phi)^2_{\rm UR}
    :=\left(\int_{t_{i}}^{t}ds\,(\delta\dot\phi(s))_{\rm UR}\right)^2
    \sim {H^2\over 9}
    \left(\Gamma H^2\right) \left({p_{min}^6(t_i)\over {\alpha'}^2H^6}\right). 
\label{uncertain}
\end{equation}
Thus we find that this effect is also small 
compared with $\langle(\delta\phi)^2\rangle_{\rm QF}\sim H^2$ if 
$t_{i}$ is set at a time well after the scale of our interest 
exceeds the Hubble horizon scale and unless $\Gamma$ is extremely large. 
Further we note that 
\begin{equation}
 {(\delta\phi)^2_{\rm UR}\over (\delta\phi)^2_{\rm WP}}
 \sim \left\{{\Gamma H^2\over 3} 
 \left({p_{min}^3(t_i)\over \alpha' H^3}\right)\right\}^2. 
\end{equation}
Hence, if $t_{i}$ is set to satisfy $p_{min}(t_i)\ll H$, 
we can also choose $(\delta\phi)^2_{\rm WP}$ to satisfy 
$(\delta\phi)^2_{\rm WP}\gg (\delta\phi)^2_{\rm UR}$. 
The restriction to the initial time obtained here 
is consistent with the general belief that the 
quantum fluctuations of the inflaton filed 
become classical only after the horizon crossing. 

\section{master equation}

In this section we directly study the evolution of the density 
matrix. 
We choose one solution of an approximate equation 
of motion of $\phi(t)$ obtained by neglecting 
the effect of the environment. 
We denote this classical trajectory as $\bar\phi(t)$. 
Namely, $\bar\phi(t)$ satisfies 
\begin{equation}
 \ddot {\bar\phi}(t)+3H\dot {\bar\phi}(t)+V'(\bar\phi(t))=0.
\end{equation}
Here we introduce new variables, $\varphi$ and $\varphi'$, 
which represent the deviations of the $\phi$ and $\phi'$ 
from the classical trajectory $\bar\phi$ by 
\begin{equation}
 \phi=\bar\phi+\varphi,\quad
 \phi'=\bar\phi+\varphi',  
\end{equation}
and assume that $\varphi$ and $\varphi'$ are small. 
Then the effective action 
$S[\phi,\phi']=iS_{\rm sys}[\phi]-iS_{\rm sys}[\phi']+
iS_{\rm IF}[\phi,\phi']$ is reduced to 
\begin{eqnarray}
 S[\phi,\phi']=&&\Omega\int_{t_{i}}^{t}ds~
  a^3(s)\Biggl\{\left({1\over 2}\dot\varphi^2(s) 
      -{1\over 2}V''(\bar\phi(s))\varphi^2(s)+O(\varphi^3)\right)
\cr &&\hspace{2cm}
  -\left({1\over 2}{\dot\varphi}'{}^2(s) 
   -{1\over 2}V''(\bar\phi(s)){\varphi'}^2(s)+O({\varphi'}^3)
    \right)\Biggr\}
\cr
 &&-\int_{t_{i}}^{t} ds~\Delta(s)f(s)
   +i\int_{t_{i}}^{t} ds~\Delta(s) \int_{t_{i}}^{s} ds'~
        \nu(s,s')\Delta(s') +S^{(\mu)}[\phi,\phi'].
\end{eqnarray}
The time evolution operator for the density matrix 
can be obtained by constructing the Hamiltonian 
recognizing $\phi$ and $\phi'$ as 
two different interacting fields. 
Neglecting the cubic or higher order terms in $\varphi$ and 
$\varphi'$, the Hamiltonian corresponding 
to this action is obtained as 
\begin{eqnarray}
 H(t)=&&{1\over \Omega a^3(t)}P_{+}(t) P_{\Delta}(t)
     +\Omega a^3(t) V''(\bar\phi(t))\varphi_{+}(t)
      \varphi_{\Delta}(t)
\cr &&+f(t)\Delta(t)   
   -\Delta(t) \int_{t_{i}}^{t} ds~
    \left[2\left(\mu_i(t,s)+\mu_a(t,s)\right)\Sigma(s)+ 
    \mu_b(t,s) \dot \Sigma(s)
   + i\nu(t,s)\Delta(s)\right], 
\end{eqnarray}
where we defined 
\begin{equation}
\varphi_+ :={\varphi+\varphi'\over 2},\quad
\varphi_{\Delta} :={\varphi-\varphi'},
\end{equation}
and 
\begin{equation}
 P_+ ={P+P'},\quad 
 P_{\Delta}={P-P'\over 2},
\end{equation}
are the conjugate momenta of 
$\varphi_+$ and $\varphi_{\Delta}$, respectively. 
We note that $P$ and $P'$ are the conjugate momenta of 
$\varphi$ and $\varphi'$. 
Since $\varphi$ and $P$ are quantum mechanical operators, 
they should be associated with hat, $\hat{}$ , but 
in order to keep the notational simplicity, we abbreviated them. 

In the above Hamiltonian, 
there appear Heisenberg operators at a past time. 
The existence of such operators is problematic 
in solving the evolution of the density matrix. 
To overcome this difficulty we replace the Heisenberg operators 
at a past time with those at present time by using the 
solution of lowest order Heisenberg equations\cite{hereditary}, 
which are given by 
\begin{eqnarray}
 \dot P & = & -\Omega a^3 V''(\bar\phi) \varphi, \quad
 \dot \varphi = {1\over \Omega a^3} P,\cr
 \dot P' & = & \Omega a^3 V''(\bar\phi) \varphi', \quad
 \dot \varphi' = -{1\over \Omega a^3} P'.
\end{eqnarray}
Approximating $V''(\bar\phi)$ by a constant $v$, 
we can solve these equations as 
\begin{equation}
 \varphi_{+}(s)=
{T}_1(s,t)\varphi_{+}(t) +{T}_2(s,t) P_{\Delta}(t), 
\quad
 \varphi_{\Delta}(s)=
{T}_1(s,t)\varphi_{\Delta}(t) +{T}_2(s,t) P_{+}(t), 
\label{Pphi}
\end{equation}
and ${T}_1(s,t)$ and ${T}_2(s,t)$, are calculated as 
\begin{eqnarray}
 T_1(s,t) & = & {\lambda_1\over D}e^{\lambda_2(t-s)}
           -{\lambda_2\over D}e^{\lambda_1(t-s)}, \cr
 T_2(s,t) & = & {1\over a^3(t)\Omega D} 
         \left(e^{\lambda_2(t-s)}
           -e^{\lambda_1(t-s)}\right),  
\end{eqnarray}
where $\lambda_1,\lambda_2$ were constants 
which were defined in 
Eq.~(\ref{lambdaDef}).  
For the later purpose, we express $\varphi_{+}(t)$ 
in an alternative way as 
\begin{equation}
\varphi_{+}(t)= O_2(t) e^{\lambda_2(t-s)}
  - O_1(t) e^{\lambda_1(t-s)}, 
\label{Pphi2}
\end{equation}
where the operators $O_1(t)$ and $O_2(t)$ are 
defined by  
\begin{eqnarray}
 O_1(t) & := & {1\over D}\left(\lambda_2 \varphi_+(t)
   +{1\over a^3(t)\Omega}P_{\Delta}(t)\right),\cr
 O_2(t) & := & {1\over D}\left(\lambda_1 \varphi_+(t)
   +{1\over a^3(t)\Omega}P_{\Delta}(t)\right).  
\end{eqnarray}

Then $f$-term, that is the term in $H$ which contains $f$, 
becomes 
\begin{equation}
 H^{(f)}=\lambda(\bar\phi +\varphi_+)\varphi_{\Delta} f(t). 
\end{equation}

The $\mu$-term is explicitly written as 
\begin{eqnarray}
H^{(\mu)}&= &{\lambda^2 a^3(t)\Omega\over 32\pi^2}
        \log\left({p_{min}(t)\over H}\right)
        \varphi_{\Delta}(t)\left(\bar\phi(t)^3
            +3\bar\phi(t)^2 \varphi_{+}(t)\right)\cr
      &&  -\lambda^2\varphi_{\Delta}(t)
        \left[\bar\phi(t)+\varphi_{+}(t)\right]  
        \int_{t_{i}}^{t} ds~\left(\mu_a(t,s)\bar\phi^2(s)+ 
           \mu_b(t,s)\bar\phi(s)\dot{\bar\phi}(s)\right) \cr
      &&    -\lambda^2\varphi_{\Delta}(t)\bar\phi(t)
       \int_{t_{i}}^{t} ds~\left(
     2\mu_a(t,s) \bar\phi(s)\varphi_{+}(s)
      +\mu_b(t,s) \bar\phi(s)\dot\varphi_{+}(s)
              +\mu_b(t,s) \dot{\bar\phi}(s)\varphi_{+}(s)\right).  
\end{eqnarray}
The first line corresponds to the instantaneous part, $\mu_i$-term. 
In the second line, the ratio between the first and second term 
in the round bracket is given by 
\begin{equation}
\left\vert {\mu_a(t,s)\bar\phi(t)\over \mu_b(t,s)\dot{\bar\phi}(t)}
\right\vert
\sim{a^2(t)\over a^2(s)}{H\bar\phi(t)\over \dot{\bar\phi}(t)}
\sim 9{a^2(t)\over a^2(s)}\left({H^2\over v}\right), 
\end{equation}
and is found to be much greater than unity. 
Thus the second term can be neglected. 
By the same reason, the last term in 
the round bracket in the last line can be also neglected. 
Then the last line is rewritten 
by using the relation (\ref{Pphi2}) as 
\begin{eqnarray}
  \to \lambda^2 \varphi_{\Delta}(t)\bar\phi(t) O_1 (t)
   \int_{t_{i}}^{t} &ds& \left\{2\mu_a(t,s) \bar\phi(s) 
   -\lambda_1 \mu_b(t,s)\bar\phi(s)\right\}e^{\lambda_1(t-s)}
\cr
  &-& \lambda^2 \varphi_{\Delta}(t)\bar\phi(t) O_2 (t)
   \int_{t_{i}}^{t}ds\left\{2\mu_a(t,s) \bar\phi(s) 
   -\lambda_2 \mu_b(t,s)\bar\phi(s)\right\}e^{\lambda_2(t-s)}. 
\end{eqnarray}
Substituting the explicit form of $\mu_a(t,s)$ and $\mu_b(t,s)$, 
and approximating 
$\bar\phi(s)$ by a constant $\bar\phi$, we will find that 
all the dominant contribution arises from the terms that contain 
$\mu_a(t,s)$. Using the formulas given in appendix B, 
we obtain 
\begin{equation}
 H^{(\mu)}=\mu_1(t)\varphi_{\Delta}\bar\phi 
          +\mu_2(t)\varphi_{\Delta}\varphi_+(t)
          +\mu_3(t)\varphi_{\Delta}O_1(t)
          +\mu_4(t)\varphi_{\Delta}O_2(t),            
\end{equation}
with 
\begin{eqnarray}
\vert \mu_1(t)\vert, 
\vert \mu_2(t)\vert,
\vert \mu_4(t)\vert 
& \alt & {\lambda^2 a^3(t)\Omega\over 16\pi^2}
     \bar\phi^2 \left(H\Delta t\right)^2, 
\cr
\vert \mu_3(t)\vert 
& \alt & {\lambda^2 a^6(t)\Omega\over 16\pi^2 a^3(t_i)}
     \bar\phi^2 \left(H\Delta t\right), 
\end{eqnarray}
where again we approximated $\bar\phi(t)$ by a constant $\bar\phi$.

The $\nu$-term can be also rewritten by using the relation (\ref{Pphi}) as 
\begin{equation}
 H_{\nu}=-i\nu_1(t) \varphi_{\Delta}^2(t)
       -i\nu_2(t) \varphi_{\Delta}(t)P_{+}(t)+O(\varphi^3). 
\end{equation}
The coefficients are evaluated as 
\begin{eqnarray}
 \nu_1(t) & \sim & {\lambda^2}\bar\phi(t) 
     \int_{t_{i}}^t ds~ \nu(t,s)
        \bar\phi(s) T_1(s,t)\cr
 &\sim & {\lambda^2\bar\phi^2\over 
     144\pi^2 p_{min}^3(t)} a^3(t) \Omega H^3
    e^{-\lambda_2(t-t_i)}
   =: e^{6Ht}e^{-\lambda_2(t-t_{i})}\tilde\nu\cr
 \nu_2(t) & \sim & {\lambda^2}\bar\phi(t) 
     \int_{t_{i}}^t ds~ \nu(t,s)
        \bar\phi(s) T_2(s,t)\cr
  & \sim & -{e^{6Ht}\over a^3(t)\Omega}
       {\tilde\nu\over\lambda_2}\left(1-e^{-\lambda_2(t-t_{i})}\right),  
\end{eqnarray}
where we used $H(t-t_{i})\gg 1$  
and $\lambda_2(t-t_{i}) = O(1)$. 
Strictly speaking, the former is not the case for $t\sim t_i$. 
However, the absolute value of the correct expression 
does not become much larger than that of this approximate 
expression. Since in the later calculation we will find that 
the contribution from $t\sim t_i$ does not become significant, 
this approximation is not so bad. 

In the above calculations, we used many crude approximations. 
But small errors caused by these 
approximations will not significantly affect on the results 
obtained by the following discussion.  

Putting all the results together, we get 
the Hamiltonian that does not contain any hereditary terms 
as 
\begin{eqnarray}
H&=&{1\over \Omega a^3(t)}P_{\Delta} P_{+} 
   +\Omega a^3(t) v(t)\varphi_{\Delta} \varphi_{+}
   +u(t)\varphi_{\Delta} \cr
 &&  +\mu_3(t)\varphi_{\Delta}(t) O_1(t)
   +\mu_4(t)\varphi_{\Delta}(t) O_2(t)
   -i\nu_1(t)\varphi_{\Delta}^2 
   -i\nu_2(t)\varphi_{\Delta}P_{+},
\end{eqnarray}
where we defined 
\begin{eqnarray}
 v(t) & := & V''(\bar\phi(t))+{1\over \Omega a^3(t)}
   \left(\lambda f(t)+\mu_2(t)\right), \cr
 u(t) & := & \lambda f(t)\bar\phi(t)+\mu_1(t)\bar\phi(t). 
\end{eqnarray}
The master equation can be derived from the above Hamiltonian.
In the coordinate representation, it becomes 
$i\partial\tilde\rho/\partial t= H\tilde \rho$, 
where $P_+$ and $P_{\Delta}$ in $H$ are to be replaced 
by $-i\displaystyle{\partial\over\partial\varphi_+}$ 
and $-i\displaystyle{\partial\over\partial\varphi_{\Delta}}$, 
respectively. 
We note that the same master equation can be derived by means of 
different methods\cite{hereditary}.
$v(t)$ is dominated by the first term due to the same reason 
explained around Eqs.~(\ref{eq413}), (\ref{comp}) and 
(\ref{eq415}) in the preceding section. 
So we approximate $v(t)$ by a constant $v$ as before. 
Later $u(t)$ is found to be compared with $a^3(t) \Omega V'$. 
Since the former is much smaller than the latter 
because of the same reason, the 
effect of $u(t)$ is negligible small. However, we keep this term 
for a while until this fact turns out to be manifest. 

In order to solve the evolution of the reduced 
density matrix $\tilde\rho(\varphi,\varphi';t)$, 
it is better to consider in the $k-\varphi_{\Delta}$ representation, 
which is defined by 
\begin{equation}
 \zeta(k,\varphi_{\Delta};t)
=\int d\varphi_{+}\,\exp\left[{-ik\varphi_{+} a^3(t)\Omega}\right] 
 \tilde\rho(\varphi_{+},\varphi_{\Delta};t).
\label{kdelta}
\end{equation}
Then the evolution equation for $\zeta$ becomes 
\begin{equation}
 {\partial\over \partial t} \zeta=
  \Biggl[3Hk{\partial\over \partial k}
         -k{\partial\over \partial \varphi_{\Delta}}
         +v\varphi_{\Delta}{\partial\over \partial k}
       -iu\varphi_{\Delta}
     +\mu_3(t)\varphi_{\Delta}\tilde O_1
     +\mu_4(t)\varphi_{\Delta}\tilde O_2 
     -\nu_1 \varphi_{\Delta}^2 
     -a^3\Omega\nu_2 k\varphi_{\Delta}\Biggr] \zeta,  
\end{equation}
where we introduced 
\begin{eqnarray}
\tilde O_1 & := & {1\over a^3(t)\Omega D}
 \left[\lambda_2{\partial\over\partial k}
         -{\partial\over\partial \varphi_{\Delta}}\right], \cr 
\tilde O_2 & := & {1\over a^3(t)\Omega D}
 \left[\lambda_1{\partial\over\partial k}
         -{\partial\over\partial \varphi_{\Delta}}\right]. 
\end{eqnarray}
 
Let us assume the Gaussian form of the density matrix as 
\begin{equation}
 \zeta={\cal N}_{\zeta}\exp\left[
  -{1\over 2}\sum_{i,j} M_{ij} x_i x_j -\sum_{i}N_i x_i\right], 
\end{equation}
where $x_i=(k,\varphi_{\Delta})$ 
and ${\cal N}_{\zeta}$ is a time independent 
normalization constant. 
Then the evolution equation for the density matrix 
reduces to the following set of equations:
\begin{eqnarray}
{d\over dt}{\bf M} & = & \left(\begin{array}{ccc}
   6H & -2 & 0  \\
   v  & 3H & -1 \\
   0  & 2v & 0  \end{array}\right){\bf M} 
+ \left(\begin{array}{c}
   0 \\
   a^3\Omega \nu_2 \\
   2\nu_1 \end{array}\right)\cr
&& \quad\quad
   -{1\over a^3(t)\Omega D} \left(\begin{array}{c}
   0 \\
   \mu_3\left(M_{k\Delta}-\lambda_2 M_{kk}\right)
   +\mu_4\left(M_{k\Delta}-\lambda_1 M_{kk}\right) \\
   2\mu_3\left(M_{\Delta\Delta}-\lambda_2 M_{k\Delta}\right)
   +2\mu_4\left(M_{\Delta\Delta}-\lambda_1 M_{k\Delta}\right)
   \end{array}\right),\cr
{d\over dt}{\bf N} & = & \left(\begin{array}{cc}
   3H & -1 \\
   v  &  0 \\\end{array}\right){\bf N} + 
  \left(\begin{array}{c}
   0 \\
   iu \end{array}\right),
\end{eqnarray}
where 
\begin{equation}
 {\bf M} = \left(\begin{array}{c}
   M_{kk} \\
   M_{k\Delta} \\
   M_{\Delta\Delta} \end{array}\right),\quad 
{\bf N} = \left(\begin{array}{c}
   N_{k} \\
   N_{\Delta} \end{array}\right). 
\end{equation}

We first consider the evolution of  $M_{ij}$. 
For this purpose, we define 
\begin{eqnarray}
 &&\sigma_1:=3H,   \quad 
   \sigma_{2,3}:=3H\pm D, \cr
 &&{\bf e}_1:=\left(\begin{array}{c}
   2 \\
   \sigma_1 \\
   2v \end{array}\right), \quad
   {\bf e}_{2}:=\left(\begin{array}{c}
   2 \\
   \sigma_{3}\\
   {\sigma_{3}^2/2}\end{array}\right),\quad
   {\bf e}_{3}:=\left(\begin{array}{c}
   2 \\
   \sigma_{2}\\
   {\sigma_{2}^2/2}\end{array}\right).
\end{eqnarray}
Then 
\begin{equation}
\left(\begin{array}{c}
   0 \\
   0 \\
   1 \end{array}\right)
={1\over 2 D^2}\left({\bf e}_2+{\bf e}_3-2{\bf e}_1\right),
 \quad
\left(\begin{array}{c}
   0 \\
   1 \\
   0 \end{array}\right)
={1\over 2 D^2}\left(2\sigma_1 {\bf e}_1
  -\sigma_2 {\bf e}_2-\sigma_3 {\bf e}_3\right), 
\end{equation}
follows. 
Introducing new parameterization of $M_{ij}$ by 
\begin{equation}
 {\bf M}=M_1 {\bf e}_1+M_2 {\bf e}_2+M_3 {\bf e}_3, 
\label{eq24}
\end{equation}
the equations for $M_{j}$, where $j=1,2,$ or $3$, 
are decoupled like 
\begin{equation}
 {dM_j\over dt}=\sigma_j M_j+ S_{Mj},
\label{eqMj}
\end{equation}
where 
\begin{equation}
 \left(\begin{array}{c}
   S_{M1} \\
   S_{M2} \\
   S_{M3} \end{array}\right)= 
 {\nu_1\over D^2}
  \left(\begin{array}{c}
   -2 \\
   1 \\
   1 \end{array}\right)
 +{a^3\Omega\nu_2\over D^2}
  \left(\begin{array}{c}
   \sigma_1 \\
   -\sigma_2 /2 \\
   -\sigma_3 /2 \end{array}\right)
 -{\mu_3\over a^3\Omega D}
  \left(\begin{array}{c}
   M_1-2M_3 \\
   -M_1 \\
   2M_3 \end{array}\right)
 -{\mu_4\over a^3\Omega D}
  \left(\begin{array}{c}
   M_1-2M_2 \\
   2M_2 \\
   -M_1 \end{array}\right). 
\label{eqSMi}
\end{equation}
Here $\mu$-terms contain $M_i$ but we can solve the above 
equation as if $S_{Mj}$ is a given source term. 
We solve 
the above equation perturbatively taking $\lambda$ as a small 
parameter. 
In this sense, $M_j$ is also expanded in powers of $\lambda$. 
At the lowest order, we solve the homogeneous equation 
without source term in Eq.~(\ref{eqMj}). 
Then to find the next order solution of Eq.~(\ref{eqMj}), 
we need to solve the equation with the source term, 
$S_{Mj}$. At this stage, we can substitute the lowest order 
solution, $M_j^{(0)}(t)$, 
into $S_{Mj}$. Then $S_{Mj}$ can be considered 
as a given source term. 
Here we should keep in mind the limitation of the present 
analysis. In deriving Eq.~(\ref{eqMj}), 
only the one loop order correction was taken into account, and 
the lowest order Heisenberg equation was used to 
remove hereditary terms in the Hamiltonian. 
Thus only the correction up to $O(\lambda^2)$ is valid. 
If we baldly solve Eq.~(\ref{eqMj}) without 
regard to this limitation, 
many unphysical pathological features will give arise.  

{}Formally, the solution is given by 
\begin{equation}
 M_j(t)=e^{\sigma_j (t-t_{i})}M_j(t_{i})
   + \delta M_j(t), 
\label{eq26}
\end{equation}
where
\begin{equation}
 \delta M_j(t)=\int_{t_{i}}^{t} ds\, e^{\sigma_j (t-s)} S_{Mj}(s).
\end{equation}

The contribution from $\nu$-terms can be evaluated 
without specifying the lowest order solution, $M^{(0)}_j(t)$. 
The corresponding inhomogeneous solution is 
approximately evaluated as 
\begin{equation}
\left(\begin{array}{c} 
   \delta M^{(\nu)}_1(t) \\
   \delta M^{(\nu)}_2(t) \\
   \delta M^{(\nu)}_3(t) \end{array}\right)
\sim
 {\nu_1(t)\over 9H^2} 
 \left(\begin{array}{c} 
   (1-e^{-\lambda_2(t-t_{i})})/\lambda_2 \\
   ({3H/ 2\lambda_2^2})(1-e^{-\lambda_2(t-t_{i})})^2  \\
   1/(6H) \end{array}\right), 
\end{equation}

Next we consider $N$-part. In the similar manner, 
we can get the solution as
\begin{equation}
\left(\begin{array}{c} 
  N_k(t)\\
  N_{\Delta}(t) \end{array}\right)
=\left(\begin{array}{cc} 
  1 & 1\\
  \lambda_2 &\lambda_1 \end{array}\right)
 \left(\begin{array}{c} 
  N_1(t)\\
  N_2(t) \end{array}\right),
\label{NkD12}
\end{equation}
with
\begin{equation}
 N_j(t)=e^{\lambda_j (t-t_{i})} N_j(t_{i})
 +\delta N_j(t), \quad (j=1,2). 
\label{Njt}
\end{equation}
Here $\delta N_j(t)$ is a inhomogeneous solution related to 
the $u$-term. 
In the later discussion, we conclude that the effect of 
these terms can be neglected. 
Thus for the present purpose, we have only  
to know their order of magnitude.  
Hence, we roughly approximate the time dependence of $u(t)$ by 
$\propto a^3(t)$. 
Then we obtain  
\begin{equation}
\left(\begin{array}{c} 
  \delta N_1(t)\\
  \delta N_2(t) \end{array}\right)
\sim 
 {i u(t)\over 9H^2} \left(\begin{array}{c}  
   -3H(1-e^{-\lambda_2(t-t_{i})})/\lambda_2 \\
   1 \end{array}\right).
\label{N12eq}
\end{equation}

Now we discuss the initial condition for 
the reduced density matrix. 
We set the initial condition at a time 
when the size of the fluctuations of our interest, 
$k_{c}^{-1}a(t)\equiv (\alpha^{-1}\Omega)^{1/3}a(t)$, becomes 
larger than the horizon scale, $H^{-1}$, where 
$\alpha=(2\pi)^{-3}$. 
In our present approximation, the earlier epoch is inaccessible, 
for we used the evaluation of the effective action, i.e., 
the coefficients $f,\mu$ and $\nu$, under 
the assumption that $a(t)H\gg k_{min}\sim k_c$. 
Here we assume that the evolution of the fluctuations 
for $t<t_{i}$ 
can be approximated by the evolution of a non interacting field.
Then the wave function will be given by 
\begin{equation}
 \Psi(\varphi)\propto \exp\left[-{A\over 2}
    \varphi^2\right], 
\end{equation}
where 
\begin{equation}
A={\alpha \over H^2}\left(1+i{H\over p_{c}(t_{i})}\right),
\end{equation}
and 
$p_c(t)=k_c/a(t)$. 

As mentioned in Sec.~2,
we decompose the wave function into a superposition of 
Gaussian wave packets 
\begin{equation}
 \Psi_{\psi}(\varphi)\propto\exp\left[-{\Gamma\over 2}
(\varphi-\psi)^2\right], 
\end{equation}
as 
\begin{equation}
 \Psi(\varphi)\propto
 \int d\psi\, e^{-{F\over 2}\psi^2} \Psi_{\psi}(\varphi), 
\end{equation}
where 
\begin{equation}
F={A\Gamma\over \Gamma-A}.
\end{equation}
As before, $\psi$ represents the initial position of the 
peak of wave packets and it is real. 
For simplicity, we set $\Gamma$ is also real. 
The peak width of the wave packet, $\Gamma^{-1/2}$ 
should be sufficiently small compared with the 
extension of the wave function, 
$\sqrt{\langle\varphi^2\rangle_{\rm QF}}\sim H$. 
Here we choose, $\Gamma$ to satisfy 
\begin{equation}
  H^2\Gamma 
  \gg {\alpha H\over p_c(t_{i})}> 1. 
\label{Sigmac}
\end{equation}
The latter inequality comes from the fact 
that the initial condition is set after the 
scale of the fluctuations of our interest becomes 
larger than the Hubble horizon sale. 
With this choice of $\Gamma$, we find that $F\sim A$. 

Further, we assume that the initial condition is set  
at a time when 
\begin{equation}
 \lambda \ll {p_{min}^3(t_{i})\over H^3} \ll 1,
\label{initlim}
\end{equation}
is satisfied. This choice of the initial time 
much simplifies the following analysis.  
Further we assume 
\begin{equation}
 \Gamma H^2\ll \lambda \left({H\over p_{min}(t_{i})}\right)^3
 \left({H\over p_{min}(t)}\right)^3.  
\label{Gammalim}
\end{equation}
We refer to this case specified by Eqs.~(\ref{initlim}) and 
(\ref{Gammalim}) as case A. 
In case A, as the restriction to $\Gamma$ is mild, 
we can examine the dependence of the evolution of the 
density matrix on the choice of $\Gamma$. 

Alternatively, instead of the limitation on the 
initial time Eq.~(\ref{initlim}), we can set a 
rather strong limitation on the initial width 
of the wave packets as 
\begin{equation}
 {1\over \lambda}\gg H^2\Gamma. 
\end{equation}
We refer to this case as case B. 

It will be possible to examine more general cases but 
the analysis becomes much more complicated. 
So here we restrict our attention to these limited two cases. 

Now the reduced density matrix is also decomposed as Eq.(\ref{redden}) 
\begin{equation}
 \tilde\rho(t)\propto \int d\psi \int d\psi'~ 
   \exp\left[-{F\over 2}\psi^2+{F^{*}\over 2}{\psi'}^2\right] 
 \rho_{\psi\psi'}(t), 
\end{equation}
where we introduced the partial reduced density matrix, 
$\rho_{\psi\psi'}(t)$, which satisfies the same evolution 
equation as $\tilde\rho(t)$ does and its amplitude 
describes the coherence between the two worlds labeled by 
the peaks of the wave packets, $\psi$ and $\psi'$. 
The initial condition of the partial density matrix is 
given by 
\begin{equation}
 \rho_{\psi\psi'}(t_{i})\propto 
 \exp\left[-{\Gamma\over 2}\left\{(\varphi-\psi)^2+(\varphi-\psi')^2 
    \right\}\right].
\end{equation}

If we introduce the notation 
\begin{equation}
 \rho_{\psi\psi'}(t)
={\cal N}\exp\left[-{1\over 2}\sum_{i,j}m_{ij} y_i y_j 
   -\sum_i n_i y_i\right], 
\end{equation}
where $y_i=(\varphi_+,\varphi_{\Delta})$. 
The coefficients $m_{ij}$ and $n_i$ are related to 
those in the corresponding $k-\Delta$ representation, 
$M_{ij}$ and $N_i$, as 
\begin{eqnarray}
 && m_{++}={a^6\Omega^2\over M_{kk}},\quad
 m_{+\Delta}=ia^3\Omega{M_{k\Delta}\over M_{kk}}, \quad
 m_{\Delta\Delta}=M_{\Delta\Delta}-{M_{k\Delta}^2\over M_{kk}},
\cr &&
 n_{+}=ia^3\Omega{N_k\over M_{kk}}, \quad
 n_{\Delta}=N_{\Delta}-{M_{k\Delta}\over M_{kk}}N_k,\quad
 {\cal N}\sim {\cal N}_{\zeta}\exp\left[-{n_+^2\over m_{++}}\right].
\end{eqnarray}

Thus we obtain 
\begin{eqnarray}
 M_{kk}(t_{i})& =& {a^6(t_{i})\Omega^2 \over 2\Gamma},\quad
 M_{k\Delta}(t_{i}) = 0,\quad
 M_{\Delta\Delta}(t_{i}) = 
   {\Gamma\over 2},\cr
 N_k(t_{i}) & = & ia^3(t_{i})\Omega \psi_{+},\quad
 N_{\Delta}(t_{i}) = {\Gamma\over 2} \psi_{\Delta},
\label{MNinit}
\end{eqnarray}
where 
\begin{equation}
\psi_+:={\psi+\psi' \over 2},\quad
\psi_{\Delta}:={\psi-\psi'},
\end{equation}

Noting that 
\begin{equation}
 \left(\begin{array}{c}
 M_1\\
 M_2\\
 M_3\end{array}\right)
={1\over D^2}\left(\begin{array}{ccc}
 -v & \sigma_1 & -1 \\
 {\sigma_2^2/ 8} & -{\sigma_2/ 2} & {1/ 2} \\
 {\sigma_3^2/ 8} & -{\sigma_3/ 2} & {1/ 2} 
 \end{array}\right)
 \left(\begin{array}{c}
 M_{kk}\\
 M_{k\Delta}\\
 M_{\Delta\Delta} \end{array}\right), \quad
 \left(\begin{array}{c}
 N_1\\
 N_2\end{array}\right)
={1\over D}\left(\begin{array}{cc}
  \lambda_1 & -1 \\
  -{\lambda_2} & {1} 
 \end{array}\right)
 \left(\begin{array}{c}
 N_{k}\\
 N_{\Delta} \end{array}\right), 
\label{mMtrans}
\end{equation}
we can calculate $M_{j}(t_{i})$ as  
\begin{equation}
\left(\begin{array}{c}
 M_1(t_{i})\\
 M_2(t_{i})\\
 M_3(t_{i})\end{array}\right)
 \sim{a^6(t_{i})\Omega^2 \over 2\Gamma D^2}
\left(\begin{array}{c}
   -v\\
  {\sigma_2^2/ 8}\\
  {\sigma_3^2/ 8}\end{array}\right)
+{\Gamma\over 2D^2}\left(\begin{array}{c}
   -1\\
  {1/2}\\
  {1/2}\end{array}\right) .
\label{Minit}
\end{equation}
Then the lowest order solution in $\lambda$ is 
given by 
$M^{(0)}_j(t)=e^{\sigma_j(t-t_{i})}M_j(t_{i})$. 
Roughly speaking, when $\Gamma$ is not so large and  
satisfies  
\begin{equation}
 H^3 a^3(t_{i})\Omega\equiv {\alpha H^3\over p_c^3(t_{i})}\gg H^2 \Gamma, 
\label{smallGamma}
\end{equation}
the first term in the right hand side of Eq.~(\ref{Minit}) 
dominates. 
In case B, this is always the case if $\lambda\gg p^3_{min}(t_{i})/H^3$. 
For the later convenience, we introduce 
\begin{equation}
\tilde\Gamma:=\left({1\over\Gamma}
 +{\Gamma\over D^2 a^6(t_{i}) \Omega^2}\right)^{-1}. 
\end{equation}
Then it follws that  $M_2(t_{i}) \sim a^6(t_{i})\Omega^2/2D^2\tilde\Gamma$. 
We note that the order of magnitude of the initial value of the 
other components is the same or smaller than that of $M_2(t_{i})$. 

To obtain the next order correction, first we need 
to evaluate the source term $S_{Mj}$. 
The order of magnitude of $\nu$-term is  
estimated as
\begin{equation} 
S_{Mj}^{(\nu)}=O\left({
  \lambda^2\bar\phi^2 a^6(t)\Omega^2 H\over 144\pi^2\alpha'}\right), 
\end{equation}
while that of $\mu$-term is estimated as 
\begin{equation}
\left(\begin{array}{c} 
    \vert S_{M1}^{(\mu)}\vert\\
    \vert S_{M2}^{(\mu)}\vert\\
    \vert S_{M3}^{(\mu)}\vert \end{array}\right) 
\lesssim 
    {\lambda^2\bar\phi^2 a^6 (t)\Omega^2 H(H\Delta t)^2 
    \over 16\pi^2 \tilde\Gamma H^2}\left(
    \begin{array}{c} 
    1\\
    1\\
    a^3(t_{i})/a^3(t)\end{array}\right). 
\end{equation}
Now it is clear that the contribution from the $\mu$-term is 
$4\alpha'(H\Delta t)^2/(\tilde\Gamma H^2)$ 
times smaller than that from the $\nu$-term. 
This factor can be set small in both cases A and B by 
choosing $\Gamma$ appropriately.
This suppression of the contribution from the $\mu$-term 
is not so trivial. The time dependence of 
$\mu_3(t)$ is approximately 
proportional to $a^6(t)$. Hence, if there appears the combination,  
$a^{-3}(t)\mu_3(t) M_2(t)$, in $S_{Mj}^{(\mu)}$, it behaves as 
$a^9(t)$ and dominates the source term when $a^3(t)/a^3(t_{i})$ becomes 
exponentially large. The disappearance of this kind of 
dangerous terms is not manifest in Eqs.~(\ref{eqMj}) and (\ref{eqSMi}). 
Here we should note that the contribution of 
$\mu$-term to $\delta M_3$, is much more suppressed 
by the existence of the factor $a^3(t_{i})/a^3(t)$ 
compared with that of the $\nu$-term. 
Thus, $\delta M_1$ and $\delta M_2$ might be dominated by the 
$\mu$-term but $\delta M_3$ is not. 

Under the conditions for case A or case B, 
$M_1(t)$ and $M_3(t)$ are dominated by the inhomogeneous solution, 
$\delta M_j(t)$, while  
the contribution to $M_2(t)$ from $\delta M_2(t)$ gives only 
a small collection to $M_2(t)$. 
Then from Eq.~(\ref{eq26}), 
we find 
\begin{eqnarray}
 M_1(t) & \sim &  {\lambda^2\bar\phi^2 a^3(t)\Omega
    \over 1296\pi^2 p_{min}^3(t)}
    {H(1-e^{-\lambda_2(t-t_{i})})\over\lambda_2},\quad\cr
 M_2(t) &\sim& e^{\sigma_2(t-t_{i})} M_2(t_{i})+\delta M_2(t) 
 \sim {a^6(t)\Omega^2 \over 4\tilde\Gamma}e^{-\sigma_3(t-t_{i})} 
    +\delta M_2(t), \quad\cr
 M_3(t) &\sim &{\lambda^2\bar\phi^2 a^3(t)\Omega
    \over 7776\pi^2 p_{min}^3(t)}. 
\label{eq37}
\end{eqnarray}
Note that they are all real. 

If we again substitute this solution into $S_{Mj}^{(\mu)}$, its 
time dependence becomes proportional to $a^9(t)$ 
while that of $S_{Mj}^{(\nu)}$ is given by $a^6(t)$. 
Thus it seems that $S_{Mj}^{(\mu)}$ becomes dominant when 
$a^3(t)/a^3(t_{i})$ becomes large. 
However, this does not mean the breakdown of the 
present perturbation scheme. 
In the present calculation, we used the lowest order 
Heisenberg equation in deriving Eq.~(\ref{eqMj}). 
Thus as mentioned before, such a substitution of $\delta M_j$ 
into $S_{Mj}^{(\mu)}$ is not allowed. 

Using Eqs.~(\ref{eq24}), (\ref{mMtrans}) 
and (\ref{eq37}), 
we finally obtain 
\begin{eqnarray}
m_{++}(t) && \sim {a^6(t)\Omega^2\over 2M_2(t)}
\sim e^{\sigma_3(t-t_{i})} {a^6(t_{i})\Omega^2\over 2M_2(t_{i})}
 \sim 2\tilde\Gamma e^{\sigma_3(t-t_{i})}
,\cr
m_{+\Delta}(t) && \sim {i v \over 3 H}a^3(t)\Omega,\cr
m_{\Delta\Delta}(t) &&=\left(4 D^2 M_2(t) M_3(t)- D^2 M_1^2(t)
           \right)/2M_2(t) 
\sim 18H^2 M_3(t). 
\end{eqnarray}
It should be mentioned that $m_{+\Delta}$ stays purely imaginary. 
Thus it does not contribute to the 
absolute magnitude of the density matrix. 

{}For $n_{j}$, with the aid of Eqs.~(\ref{Njt}), (\ref{MNinit}) 
and (\ref{mMtrans}), we obtain 
\begin{eqnarray}
n_{+}(t) && \sim ia^3(t)\Omega{N_1(t)\over 2M_2(t)}\cr
&&\sim 2\tilde\Gamma\left[
 -e^{\lambda_2(t-t_{i})}\psi_+ +e^{\sigma_3(t-t_{i})}
  \left({u(t)\over 3H^2a^3(t)\Omega}\right) 
   {H(1-e^{-\lambda_2(t-t_{i})})\over \lambda_2} 
 - ie^{\lambda_2(t-t_{i})}{\Gamma\psi_{\Delta}\over 
    6H a^3(t_{i})\Omega}\right],\cr
n_{\Delta}(t) && ={D\over 2M_{2}(t)}\left(2 M_2(t) N_2(t)- 2M_3(t) N_1(t) 
           + M_1(t)(N_2(t) -N_1(t))\right)\cr
&& \sim -i\left[-{u(t)\over 3H}+e^{\lambda_2(t-t_{i})} 
  {12H\tilde\Gamma M_1(t)\over a^3(t)\Omega}\psi_{+}\right]
  +e^{\lambda_2(t-t_{i})} 
   {2\Gamma\tilde\Gamma M_1(t)\over a^3(t) a^3(t_{i}) \Omega^2}
   \psi_{\Delta}. 
\end{eqnarray}
Here we note that in the above evaluations of $m_{ij}(t)$ and $n_j(t)$ 
there is no relevant contribution from $\delta M_1$ and 
$\delta M_2$. Among $\delta M_j$ 
a relevant contribution is provided only by 
$\delta M_3$, which is almost insensitive to the 
effect of $\mu$-term. 

To see the degree of decoherence, 
we consider the following ratio 
\begin{equation}
R:={\hbox{\rm max}|\rho_{\psi\psi'}(t)|\over 
 \hbox{\rm max}|\rho_{\psi\psi'}(t_{i})|},  
\end{equation}
where max means the maximum value when   
$\varphi_+$ and $\varphi_{\Delta}$ are varied.
As was discussed in Sec.~2, 
$R$ is the quantity that represents how efficiently 
the coherence between the two worlds (two wave packets) 
labeled by, $\psi$ and $\psi'$, gets lost. 
With this definition, $R$ is evaluated by  
\begin{equation}
 R= \exp\left(K(t)-K(t_{i})\right),  
\end{equation}
with 
\begin{equation}
 K:={1\over 2}\left({\Re(n_{\Delta})^2\over 
    m_{\Delta\Delta}}+{\Im(n_{+})^2\over 
    m_{++}}\right). 
\label{defK}
\end{equation}
In deriving this formula, we used the fact 
that ${\cal N}_{\zeta}$ is a constant and we 
neglected the small 
logarithmic correction that comes from the 
Gaussian-integral in Eq.~(\ref{kdelta}). 
Then, after a straight forward calculation, we obtain
\begin{equation}
 K(t_{i})={\Gamma\over 4}\psi_{\Delta}^2,
\end{equation}
and
\begin{equation}
 K(t)\sim \left({M_1^2(t)\over M_2(t)M_3(t)}+{1}\right)
 {\Gamma^2\tilde\Gamma\over 32 H^2 a^6(t_{i})\Omega^2 }\psi_{\Delta}^2,
\label{Kt}
\end{equation}
for $H(t-t_{i})\gg 1$. 
In Eq.~(\ref{Kt}), the first and the second 
terms in the round bracket represent the 
contribution from the corresponding 
terms in Eq.~(\ref{defK}), respectively.  
It is clear that the second term dominates $K(t)$. 

When the first term in the defining equation 
of $\tilde \Gamma$ dominates, 
i.e., when Eq.~(\ref{smallGamma}) is satisfied, 
\begin{equation}
 {\Gamma\tilde\Gamma\over 9H^2 a^6(t_{i})\Omega^2} \sim 
 {\Gamma^2\over 9H^2 a^6(t_{i})\Omega^2}\ll 1. 
\end{equation}
Hence, $K(t)$ is much smaller than $K(t_{i})$, and 
$R$ is predominantly determined by $K(t_{i})$ as 
\begin{equation}
 R\sim \exp\left(-K(t_{i})\right).
\label{Rest}
\end{equation}
This means that the coherence between two wave packets 
labeled by $\psi$ and $\psi'$ is exponentially 
suppressed for large $\psi_{\Delta}$
and the typical scale of decoherence is 
determined by the width of the wave packets. 
For two wave packets with $\psi_{\Delta}^2<\Gamma^{-1}$, 
their overlap is large. 
Hence, it is natural that their coherence 
is maintained. 
So the scale of the decoherence depends totally on 
the width of wave packets that we choose. 

In case A, we can set the width of wave packet much smaller so that 
the second term in the defining equation of $\tilde \Gamma$ dominates. 
Then 
\begin{equation}
 {\Gamma\tilde\Gamma\over 9H^2 a^6(t_{i})\Omega^2} \sim 
 1-{9H^2 a^6(t_{i})\Omega^2\over \Gamma^2}, 
\end{equation}
and $R$ becomes 
\begin{equation}
 R\sim \exp\left(-{9H^2 a^6(t_{i})\Omega^2 \over 4\Gamma}\psi_{\Delta}^2\right). 
\end{equation}
In this case 
the typical scale of the decoherence between two different wave packets, 
$(\delta\psi)^2_{\rm dec}$, will be given by 
\begin{equation}
 (\delta\psi)^2_{\rm dec}\sim {2\Gamma\over 9H^2a^6(t_{i})\Omega^2}. 
\end{equation}
Thus the decoherence does not occurs if we set the initial 
wave packets too narrow, i.e, if $\Gamma$ is too large.  
The best choice of $\Gamma$ that minimizes $R$ is given by 
\begin{equation}
 \Gamma H^2={3H^3a^3(t_i)\Omega={3}\alpha H^3\over p_c^3(t_{i})}.
\label{Gammamin}
\end{equation} 
With this choice of $\Gamma$, 
the typical scale of the decoherence becomes
\begin{equation}
 (\delta\psi)^2_{\rm dec}\sim {4 H^2\over 3} 
\left({p_{c}^3(t_{i})\over \alpha H^3}\right). 
\label{delpsidec}
\end{equation}

This bound mainly comes from the broadening of the 
wave packet due to the uncertainty relation. 
This fact can be understood by seeing that 
the minimum value of $(\delta\psi)^2_{\rm dec}$ 
correspond to the scale given in Eq.~(\ref{uncertain}) 
in the preceding section. 
To see this fact in the present context, 
we consider the expectation value 
of, $(\varphi_{+})^2$, which is expected to represent 
the degree of the broadening of the wave packet.
If $H^2\Gamma \ll {\alpha H^3/p_c^3(t_{i})}$, we have 
\begin{equation}
\langle(\varphi_+)^2\rangle \sim 2/m_{++}
\sim 1/\tilde \Gamma ,
\end{equation}
and is found to stay almost constant.  
Instead, if we assume a very large value of $\Gamma$ such that violates 
this condition, 
the broadening of the wave packet at a late time 
is evaluated as 
\begin{equation}
\langle(\varphi_+)^2\rangle \sim
 {1\over \Gamma}
   \left[{\Gamma H^2\over 3 \alpha H^3/p_c^3(t_{i})}\right]^2,
\label{broadening}
\end{equation}
and becomes much larger than the initial width of the 
wave packet $\Gamma^{-1}$. 
It is easy to see that 
the minimum broadening of the wave packet 
\begin{equation}
\langle(\varphi_+)^2\rangle \sim {2H^2\over {3}}
\left({p_c^3(t_{i})\over \alpha H^3}\right), 
\end{equation}
is also achieved when Eq.~(\ref{Gammamin}) holds. 
Comparing this result with Eq.~(\ref{delpsidec}), 
we find that the scale of decoherence is determined 
by the broadening of the wave packets. 

However, there is another broadening mechanism 
due to the $\nu$-term. 
The fluctuations of the environment behave as 
a stochastic noise, and cause the broadening of the wave packet.  
In the last section we estimated the fluctuation in $\delta\phi$ 
responsible for this effect in 
Eq.~(\ref{EAdphi}), and
we left the task to justify our 
interpretation of $\langle(\delta\phi)^2\rangle_{\rm EA}$.  
Now we return to this issue. 
In evaluating Eq.~(\ref{broadening}), we have completely 
neglected the contribution from the $\nu$-term. 
As mentioned before $M_2(t)$ has a small correction, 
$\delta M_2(t)$, due to the effect of $\nu$-term. 
This gives an additional broadening of the wave packet;
\begin{equation}
\langle\delta(\varphi_+)^2\rangle \sim 
{2\,\delta M_2(t_{i})\over a^6(t_{i})\Omega^2}
   \sim {\lambda\over 216\alpha'\pi^2} \left({H(
    1-e^{-\lambda_2(t-t_{i})})\over \lambda_2}\right)^2 
   \left({v\over H^2}\right) H^2,  
\end{equation}
which corresponds to the expression 
previously derived in Eq.~(\ref{EAdphi}) with the aid of 
the augsiliary field and naive approximations. 
At this point, the meaning of the quantity 
$\langle(\delta\phi)^2\rangle_{\rm EA}$ became transparent. 
$\langle(\delta\phi)^2\rangle_{\rm EA}$ represents the fluctuation 
caused by the environment. 
Both in case A and in case B, this broadening effect does not change 
the width of wave packets much 
compared with the case in which this effect is neglected. 
This is simply because we restricted our attention to 
the case when $M_2^{(0)}(t)\gg \delta M_2(t)$ holds 
for simplicity. 

Then we find that the typical scale of the decoherence 
is not directly related with $(\delta\phi)_{\rm dec}^2$, which 
was evaluated in Eq.~(\ref{phidec}). 
We find this scale in 
$
 M_{\Delta\Delta}^{-1},  
$
which gives exactly the same expression as $(\delta\phi)_{\rm dec}^2$. 
To understand the meaning of this result, 
we focus on one diagonal component 
of the partial reduced density 
matrix, $\rho_{\psi\psi}(\varphi,\varphi';t)$.  
$\rho_{\psi\psi}(\varphi,\varphi';t)$ also has 
two continuous arguments, $\varphi$ and $\varphi'$. 
The suppression of the off-diagonal elements of 
$\rho_{\psi\psi}(\varphi,\varphi';t)$ is determined by $m_{\Delta\Delta}$, 
and $\rho_{\psi\psi}(\varphi,\varphi';t)$ becomes exponentially small  
for a large $\varphi_{\Delta}$ which satisfies 
$\varphi_{\Delta}\gg (\delta\phi)_{\rm dec}$. 
Thus we can say that the decoherence occurs within a wave packet 
on the scale of $(\delta\phi)_{\rm dec}$. 
However, for narrower wave packets, the classicality of the 
evolution of the system cannot be maintained through the whole 
duration of our concern. In the present case, 
this condition of the classicality 
of the evolution of the system totally determines the minimum 
width of the wave packet. 

{}Finally we observe the peak location. 
It is calculated as 
\begin{equation}
 \langle \varphi_+(t) \rangle =-{\Re(n_+)\over m_{++}}
\sim e^{-\lambda_2(t-t_{i})}\psi_+ 
 +\left({u(t)\over 3H^2 a^3(t)\Omega}\right) {H(
    1-e^{-\lambda_2(t-t_{i})})\over \lambda_2}. 
\end{equation}
The first term represents the change of the separation 
of the different trajectories 
labeled by the initial position of the peak, $\psi_+$. 
This is just the term to be attributed to the nature of the 
model potential. In the present model, congruence of the 
classical trajectories converges gradually. 
The second term is independent of $\psi_+$. 
This term arises because 
the interaction with the environment was not 
taken into account when we determine $\bar\phi(t)$. 
Thus it can be interpreted as the correction to $\bar\phi(t)$ 
due to the effect of the environment. 
This correction does not change the motion of $\bar\phi(t)$ 
so much by the same reason that discussed around 
Eqs.~(\ref{eq413}), (\ref{comp}) and (\ref{eq415}). 

\section{summary and discussion}
We investigated the evolution of the perturbation of the 
inflaton field with $\lambda\phi^4$ potential 
after the scale of the perturbation 
exceeds the Hubble horizon scale during the inflation. 
The effect of the coupling to the smaller scale modes 
through the self interaction was taken into consideration 
by using the closed time path formalism. 
That is, the smaller scale modes are considered as the environment. 
The initial condition for the quantum state of the fluctuation 
of the inflaton field was set well after the time of 
the horizon crossing of the considered mode. 
The initial state was supposed to be given by a pure state 
density matrix 
composed of a direct product 
of the usual Euclidean vacuum state, 
which has the variance of $O(H^2)$. 
This initial quantum state can be recognized as a quantum mechanical 
superposition of wave packets with a sharp peek. 
In the present model, we found that the influence 
of the environment does not distort these wave packets much 
but it extinguishes the coherence between the wave packets with 
different peek positions. 
The efficiency of the decoherence is so high that 
the state described by the different wave packets 
can be recognized as completely different worlds. 
Hence, we can conclude that the initial pure state 
evolves into a mixed state which can 
be interpreted as a statistical ensemble.

In the context of the inflationary 
universe scenario, the primordial fluctuations of the universe 
are evaluated by using an ad-hoc classicalization ansatz 
such that 
the expectation value of the squared field operator can be 
interpreted as the amplitude of the variance 
of the statistical ensemble. 
In this paper we have shown that this ansatz can be verified 
in a simple model. 
Thus the result obtained here gives a 
partial justification of the standard 
calculation of the primordial fluctuations. 

However, the important issue might be whether the 
inflaton field behaves as classical during the reheating 
process that successively occurs after inflation. 
This is because the previous study on reheating is mostly 
based on the assumption that the fluctuations have already 
become classical before the reheating occurs. 
Thus, it will not be necessary that 
the fluctuations of the inflaton field are kept to 
be classical during the inflation. 
In order to prove that the standard calculation works, 
we have only to show that the fluctuations of the inflaton 
become classical not throughout the 
whole duration of the inflation but at the end of it. 
In this sense, the condition for the standard calculation 
to be justified might be weaker than the conditions required 
in the present paper. 

In this paper, we restricted our consideration to a specific model and 
the effect of the metric perturbation and the process of 
reheating were not taken into account at all. 
So further study is still required as future work. 

\acknowledgments
We thank Misao Sasaki for fruitful conversations. 
This work was supported in part 
by Monbusho Grant-in-Aid for Scientific Research No.~07304033.

\appendix
\section{Evaluation of $f(s)$, $\mu(s,s')$ and $\nu(s,s')$}
In this appendix, we show the details of the calculation 
to obtain the approximate expression for $f(s)$, $\mu(s,s')$ 
and $\nu(s,s')$ given in Eqs.~(\ref{fmunu}) and (\ref{musdef}). 

\subsection{$f(s)$}
We begin with the evaluation of $f(s)$. 
The explicit expression for $f(s)$  is written down as  
\begin{equation}
f(s)={a(s)\Omega\over 4\pi^2}\int_{k_{min}}^{\infty} 
dk{k\over 2}\left[1+{1\over k^2\eta^2}\right]. 
\end{equation}
This expression is divergent, and needs renormalization. 
For this purpose, we use the point splitting technique. 
The function, $f(s)$, is basically given by using the 
Weightman function as
$f(s)\propto\int d^3x G^{(+)}(x,s;x,s)$. 
Now we regularize the expression as $\int d^3x G^{(+)}(x,s,x',s)$, 
where $x'$ is chosen to satisfy $|x-x'|^2=z^2$. 
After this regularization, $f(s)$ is calculated straightforwardly as
\begin{eqnarray}
 f(s) & = & {a^3(s)\Omega\over 16\pi^2}
 \int_{p_{min}(s)}^{\infty} dp\int_{-1}^{1} d(\cos \theta)\, p
  \left(1+{H^2\over p^2}\right) e^{ip\cos\theta z}\cr
 & = & 
 {a^3(s)\Omega\over 16\pi^2}
 \left\{{2\over z^2} -p_{min}^2(s)+2 H^2-H^2 \Bigl[
 \hbox{\rm Ei}(-ip_{min}(s) z)+\hbox{\rm Ei}(ip_{min}(s) z)
     \Bigr]+O(z)\right\}\cr
 & \sim & {a^3(s)\Omega\over 8\pi^2}
   \left\{{1\over z^2}- H^2\log z - H^2(\gamma-1) 
    -{p^2_{min}(s)\over 2}-H^2\log(p_{min}(s))+O(z)\right\}, 
\end{eqnarray}
where $\hbox{Ei}$ is the exponential integral function. 
The divergent first term in the last line 
exists even in the limiting case where 
the background curvature can be neglected. 
Hence, it should be subtracted by the renormalization procedure. 
The second term is to be attributed to the renormalization of the 
curvature coupling term, $\xi R\phi^2$, where $R$ is 
the scalar curvature of the spacetime. 
Setting an appropriate renormalization condition, we obtain the 
renormalized $f(s)$ as 
\begin{equation}
f(s)={a^3(s)\Omega\over 8\pi^2}\left[-{1\over 2} p_{min}^2(s) 
  +H^2\log\left({H \over p_{min}(s)}\right)\right]. 
\end{equation}

\subsection{$\mu(s,s')$}
The function $\mu(s,s')$ is explicitly written down as 
\begin{eqnarray}
\mu(s,s') & = & -i{a^3(s) a^3(s')\Omega\over 4\pi^2} \int_{k_{min}}^{\infty}
  dk k^2 
   \left(u^{*}(s)^2 u(s')^2 -u(s)^2 u^{*}(s')^2\right)\cr
  &=& -i{a(s) a(s')\Omega\over 16\pi^2} 
   \int_{k_{min}}^{\infty}dk
   \left(e^{2ik(\eta-\eta')}\left(1+{i\over k\eta}\right)^2
   \left(1-{i\over k\eta'}\right)^2-(\hbox{c.c.})\right). 
\end{eqnarray}
where $\eta$ and $\eta'$ is the conformal time 
corresponding to $s$ and $s'$. 
We divide it into two pieces, 
$\mu_r(s,s')$ and  $\mu_s(s,s')$.  
\begin{equation}
\mu_s(s,s'):=
   -i{a(s) a(s')\Omega\over 16\pi^2} 
   \int_{k_{min}}^{\infty}
   \left(e^{2ik(\eta-\eta')}-(\hbox{c.c.})\right), 
\end{equation}
 is the portion that contains the ultraviolet divergence 
corresponding to coupling constant renormalization and 
$\mu_r(s,s')$ is the remaining 
regular terms defined by $\mu_r(s,s'):=\mu(s,s')-\mu_s(s,s')$. 

The expression for $\mu_r(s,s')$ is slightly 
complicated  
but there is no technical difficulty in 
its evaluation. 
After a straightforward calculation, 
we obtain 
\begin{equation}
 \mu_r(s,s') \sim 
 {a(s) a(s')\Omega\over \pi^2}\left\{
 {1\over 12}\left({1\over \eta}-{1\over \eta'}\right)
 +{1\over 12}\left[{\eta'\over \eta^2}-{\eta\over \eta'^2}\right]
 \left(-{7\over 3}+\gamma
  +\log\left[2k_{min}(\eta-\eta')\right]\right)\right\}. 
\end{equation}
Here we take into account the 
fact (\ref{etacondi}) and pick up only the dominant terms 
in the $k_{min}\rightarrow 0$ limit. 
However, this expression is still complicated.  
In this paper, we just aim to show 
that the effect of $\mu-$term is negligible small. 
For this purpose, we can simplify the expression by taking 
the $\vert\eta\vert\ll\vert\eta'\vert$ limit as 
\begin{equation}
 \mu_r(s,s')\sim-{a^3(s)\Omega H\over 12\pi^2}
   \log\left(k_{min}(\eta-\eta')\right).
\end{equation}  
When $s\sim s'$, this simplified expression is not correct. 
However, since $\mu(s,s')$ vanishes in the 
coincidence limit, this simplification does not underestimate 
the effect of $\mu-$term. 
Hence, this simplification will be justified. 

The singular part $\mu_s(s,s')$ needs regularization as before. 
Recalling that $\mu(s,s')$ is essentially given by 
the product of the Weightman function as
\begin{equation}
 \mu(s,s')\propto \int d^3 x \left[G^{(+)}(x,t;0,t')\right]^2+(\hbox{c.c.}), 
\end{equation}
we can introduce the point splitting regularization by replacing 
$\left[G^{(+)}(x,t;0,t')\right]^2$ by $G^{(+)}(x,t;0,t')
G^{(+)}(x+\epsilon,t;0,t')$, where $\vert\epsilon\vert^2=z^2$.  
Then the regularized expression for $\mu_s(s,s')$ becomes 
\begin{equation}
 \mu_s(s,s')=-i{a(s)a(s')\Omega\over 32\pi^2} 
 \int_{k_{min}}^{\infty}dk
  \int_{-1}^{1} d(\cos\theta) 
   \left(e^{2ik(\eta-\eta')}-e^{-2ik(\eta-\eta')}\right)
   e^{-ikz\cos\theta /a(s)}. 
\end{equation}

In order to subtract the singular term which needs renormalization 
it is necessary to consider the expression including the $s'$ integral
\begin{equation}
 I:=\int_{t_{i}}^{s} ds' \Sigma(s') \mu_s(s,s').
\end{equation}
Integrating by part, this integral is rewritten as 
\begin{equation}
I=\left[ a^2(s') U(s,s') \Sigma(s')\right]_{t_{i}}^{s} 
- -\int_{t_{i}}^{s} ds' U(s,s') 
   {d\left(\Sigma(s')a^2(s')\right)\over ds'}, 
\end{equation}
where 
\begin{equation}
U(s,s'):= {a(s)\Omega\over 64\pi^2}\int_{k_{min}}^{\infty} 
{dk\over k} 
\int_{-1}^{1}d(\cos\theta) 
\left[
  e^{2ik(\eta-\eta')}+e^{-2ik(\eta-\eta')}\right] 
  e^{-ik\cos\theta z/a(s)} = 
 \int^{s'} {ds''\over a^2(s'')} \mu_s(s,s'').
\end{equation}

In evaluating $U(s,s')$, when $s\ne s'$, we can take the 
$z\to 0$ limit without any trouble, and easily evaluated as 
\begin{equation}
 U(s,s')\sim -{a(s)\Omega \over 16\pi^2}
 \left[\log\left(2k_{min}(\eta-\eta')\right)+\gamma\right]. 
\end{equation}
On the other hand, when $s=s'$, $U(s,s)$ becomes 
\begin{equation}
 U(s,s)\sim {a(s)\Omega\over 16\pi^2} 
 \left[(1-\gamma-\log z)-\log (p_{min}(s))\right], 
\end{equation}
which contains the logarithmic divergence corresponding to 
the coupling constant renormalization. 
After the renormalization, we obtain 
\begin{equation}
 U(s,s)\sim -{a(s)\Omega\over 16\pi^2} 
 \log\left({p_{min}(s)\over H}\right).  
\end{equation}
Combining all the results, finally we get the expression 
given in Eq.~(\ref{SIF}). 

\subsection{$\nu(s,s')$}
In the same way, the dominant terms in  
$\nu(s,s')$ are evaluated as 
\begin{eqnarray}
&&{a(s)^3 a(s')^3\Omega\over 8\pi^2}
   \int_{k_{min}}^{k_{max}} k^2 dk 
\left( u^{*}(s)^2 u(t')^2 +u(s)^2 u^{*}(t')^2\right)
\cr
&&\quad \sim
 {a(s)a(s')\Omega\over 32\pi^2} 
\left({2\over 3}{1\over \eta^2{\eta'}^2 k_{min}^3}
 +{\sin\left[2k_{max}(\eta-\eta')\right]\over \eta-\eta'}\right). 
\end{eqnarray}
The second term becomes proportional to $\delta(\eta-\eta')$ 
in the $k_{max}\rightarrow \infty$ limit. 
Since the time dependence of the second term 
is different from that of the first one, we cannot simply 
discard the second term.  
However, the important quantity for the discussion in  
the present paper is the integral over $s'$ of the 
product of $\nu(s,s')$ and some function $F(s,s')$ which
is always a smooth function with respect to $s-s'$.  
Thus we can conclude that 
the first term gives the dominant contribution, and 
the second term can be neglected. 

\section{approximate formulas for integrals}

Here we explain the details of the approximation 
used in evaluating several integrals. 
In this appendix, we set 
$\eta=-e^{-Hs}$, $\eta'=-e^{-Hs'}$ 
and $\eta_i=-e^{-Ht_{i}}$.  

First we evaluate the following integral 
\begin{eqnarray}
 I_1 & := & \int_{t_{i}}^s ds'\, H
    \log\left(k_{min}(\eta-\eta')\right)\cr
 & = & -\int_{\eta_i}^{\eta} {d\eta'\over \eta'}
    \log\left(k_{min}(\eta-\eta')\right). 
\end{eqnarray}
Changing the variable into $x:=(\eta'-\eta)/\eta$, 
the integral becomes 
\begin{equation}
 I_1= \int_0^{(\eta_i-\eta)/\eta}{dx\over 1+x} \log x +
      \int_0^{(\eta_i-\eta)/\eta}{dx\over 1+x} \log 
     (-k_{min}\eta). 
\end{equation}
The second term is explicitly calculated to become  
$H(s-t_i)\log(-k_{min} \eta)$. 
When $(\eta_i-\eta)/\eta<1$ the first term is 
bounded by 
\begin{equation}
  \left\vert\int_0^{(\eta_i-\eta)/\eta}{dx\over 1+x} \log x
  \right\vert 
   < \left\vert
   \int_0^{1}{dx\over 1+x} \log x \right\vert 
   ={\pi^2\over 12}. 
\end{equation}
To the contrary, when $(\eta_i-\eta)/\eta>1$ the first term is 
evaluated by  
\begin{equation}
  \int_0^{(\eta_i-\eta)/\eta}{dx\over 1+x} \log x 
 =-{\pi^2\over 12}+ 
   \int_1^{(\eta_i-\eta)/\eta}{dx\over 1+x} \log x 
\end{equation}
and 
\begin{equation}
 \left\vert\int_1^{(\eta_i-\eta)/\eta}{dx\over 1+x} \log x
 \right\vert < \log{\eta_i-\eta\over \eta}
   \int_1^{(\eta_i-\eta)/\eta}{dx\over 1+x}
 = \log{\eta_i-\eta\over \eta}\log{\eta_i\over 2\eta}. 
\end{equation}
Thus we find 
\begin{equation}
    \vert I_1\vert\alt (H\Delta t)^2. 
\end{equation}

Next we consider 
\begin{eqnarray}
  I_2&:= & \int_{t_{i}}^s ds'\, H{a^2(s')\over a^2(s)}
    \log\left(k_{min}(\eta-\eta')\right)\cr
 & = &- \eta^2\int_{\eta_i}^{\eta} {d\eta'\over {\eta'}^3}
    \log\left(k_{min}(\eta-\eta')\right). 
\end{eqnarray}
After a straightforward calculation by using 
the same change of variable as before, we obtain 
\begin{equation}
  I_2= -{1\over 2}\left[{\eta^2\over \eta_i^2}
    \log\left({\eta_i\over \eta}-1\right) +
   \log\left({\eta_i\over \eta_i-\eta}\right) 
   +2{\eta-\eta_i\over \eta_i}\right] 
   +{1\over 2}\left(1-{\eta^2\over\eta_i^2}\right) 
   \log\left(-k_{min}\eta \right). 
\end{equation}
Thus we find  
\begin{equation}
 \vert I_2\vert \alt H\Delta t.
\end{equation}

Finally we consider 
\begin{eqnarray}
  I_3&:= & \int_{t_{i}}^s ds'\, H{a^3(t_{i})\over a^3(s')}
    \log\left(k_{min}(\eta-\eta')\right)\cr
 & = &{\eta_i}^{-3}\int_{\eta_i}^{\eta} {d\eta'{\eta'}^3}
    \log\left(k_{min}(\eta-\eta')\right). 
\end{eqnarray}
In the same way, 
after a straightforward calculation, 
we obtain 
\begin{equation}
  I_3= {\eta^3\over \eta_i^3}\left[
   {1\over 3}\left({\eta_i^3\over \eta^3}-1\right)
    \log\left({\eta_i\over \eta}-1\right) -
   {1\over 9}\left({\eta_i\over \eta}-1\right)^3 
   -{1\over 2} \left({\eta_i\over \eta}-1\right)^2 
   -\left({\eta_i\over \eta}-1\right)
   +{1\over 3}\left({\eta_i^3\over\eta^3}-1\right) 
   \log\left(-k_{min}\eta \right)\right], 
\end{equation}
and it implies 
\begin{equation}
 \vert I_3\vert \alt H\Delta t.
\end{equation}  

\end{document}